\documentclass{aastex}		% [manuscript] default; 1-col, 2-space

\usepackage[numberedappendix]{emulateapj5}  % manuscript mode, emulate ApJ

\newcommand{\hi}{\hbox{H~I}}
\newcommand{\hii}{\hbox{H~II}}
\newcommand{\lsun}{\hbox{$L_{\odot,B}$}}
\newcommand{\msun}{\hbox{$M_{\odot}$}}
\newcommand{\vsun}{\hbox{$v_{\odot}$}}

\newcommand{\otwo}{\hbox{[O II]$\lambda 3727$}}

\newcommand{\hgamma}{\hbox{H$\gamma$}}
\newcommand{\othreea}{\hbox{[O III]$\lambda 4363$}}
\newcommand{\hbeta}{\hbox{H$\beta$}}
\newcommand{\othree}{\hbox{[O III]$\lambda\lambda 4959,5007$}}
\newcommand{\othreeb}{\hbox{[O III]$\lambda 4959$}}
\newcommand{\othreec}{\hbox{[O III]$\lambda 5007$}}

\newcommand{\ntwotemp}{\hbox{[N II]$\lambda 5755$}}

\newcommand{\oonea}{\hbox{[O I]$\lambda 6300$}}

\newcommand{\ooneb}{\hbox{[O I]$\lambda 6364$}}

\newcommand{\halpha}{\hbox{H$\alpha$}}
\newcommand{\ntwob}{\hbox{[N II]$\lambda 6583$}}

\newcommand{\stwo}{\hbox{[S II]$\lambda\lambda 6716,6731$}}

\newcommand{\ntwootwo}{\hbox{[N II]/[O II]}}

\journalinfo{A.J., accepted (scheduled June 2003)} % emulateapj5 ONLY; UppLeft

\shortauthors{Lee, McCall, \& Richer}
\shorttitle{Clues to Galaxy Evolution. II.} 
\slugcomment{Full version with included figures at 
{\tt http://www.mpia.de/homes/lee/paper2.ps.gz}
}

\begin{document}

\title{
Uncovering Additional Clues to Galaxy Evolution. \\
II.  The Environmental Impact of the Virgo Cluster on the \\
Evolution of Dwarf Irregular Galaxies 
}

\author{Henry Lee$\,$\altaffilmark{1,2,3,4}, 
Marshall L. McCall$\,$\altaffilmark{2}, and
Michael G. Richer$\,$\altaffilmark{3,4,5}
}

\altaffiltext{1}{
Max-Planck-Institut f\"ur Astronomie, 
K\"onigstuhl 17, D-69117, Heidelberg, Germany.
E-mail: {\tt lee@mpia.de}
}

\altaffiltext{2}{Dept. of Physics \& Astronomy, York University,
4700 Keele St., Toronto, Ontario  M3J 1P3  Canada.
}

\altaffiltext{3}{Visiting Astronomer, Kitt Peak National Observatory, 
National Optical Astronomy Observatory, which is operated by the
Association of Universities for Research in Astronomy, Inc. (AURA)
under cooperative agreement with the National Science Foundation.}

\altaffiltext{4}{Visiting Astronomer, Canada-France-Hawaii Telescope, 
operated by the National Research Council of Canada, the
Centre National de la Recherche Scientifique de France, 
and the University of Hawaii.}

\altaffiltext{5}{Observatorio Astron\'omico Nacional, 
P.O. Box 439027, San Diego, CA, 92143--9027 USA.
}

\begin{abstract}		% ABSTRACT
The impact of the cluster environment on the evolution of dwarf
galaxies is investigated by comparing the properties of a sample of
dwarf irregulars (dIs) in the Virgo Cluster with a control sample of
nearby (``field'') dIs having oxygen abundances derived
from \othreea\ measurements and measured distances from resolved
stellar constituents.
Spectroscopic data are obtained for \hii\ regions in 11 Virgo 
dIs distributed in the central and outer regions of the cluster.  
To ensure that oxygen abundances are derived in a homogeneous manner,
oxygen abundances for field and Virgo dIs are computed
using the bright-line method and compared with abundances directly
obtained from \othreea, where available. 
They are found to agree to within about 0.2~dex with no systematic
offset.
At a given optical luminosity, there is no systematic difference in
oxygen abundance between the sample of Virgo dIs and the sample 
of nearby dIs. 
However, five of the eleven Virgo dIs exhibit much lower 
baryonic gas fractions than field dIs at comparable oxygen abundances.
Using field dIs as a reference, a gas-deficiency index for dIs is
constructed, making it possible quantitatively to identify which
galaxies have lost gas. 
For the Virgo sample, some of the dwarfs are gas-deficient by a factor
of 30.
The gas-deficiency correlates roughly with the X-ray surface
brightness of the intracluster gas.
Ram-pressure stripping can best explain the observed gas-poor 
dIs in the cluster sample. 
Together with the lack of significant fading and reddening of the
gas-poor dIs compared to gas-normal dIs, these observations suggest
that the gas-poor dIs in Virgo have recently encountered the
intracluster medium for the first time.
Faded remnants of gas-poor dIs in Virgo will resemble bright
dwarf ellipticals presently seen in the cluster core. 
\end{abstract}


\section{Introduction}		% INTRODUCTION
\label{sec_intro}

Interactions between galaxies and their environments may be important
factors in galaxy evolution. 
An important venue to test the impact of environmental conditions 
is within clusters of galaxies (e.g., \citealp{donahue98,bahcall99}),
where huge numbers of galaxies of various sizes, luminosities, and 
morphologies, as well as a large mass of gas, are confined within a
specific volume of space.
Indeed, numerous populations of dwarf galaxies have been discovered in
groups and clusters,
e.g., Virgo Cluster: \cite{phillipps98};
Fornax Cluster: \cite{drinkwater01};
also, see \cite{zm98,zm00}.
Various kinds of processes may act on a cluster galaxy: 
tidal forces from galaxy-galaxy and galaxy-cluster interactions
(e.g., \citealp{bv90,hb96}),
ram-pressure effects due to the intracluster medium 
(e.g., \citealp{mb2000,quilis00}), 
high-speed encounters between galaxies (e.g., \citealp{moore96}),
collisions and mergers (e.g., \citealp{bh91}),
or some combination of the above.

Owing to its proximity (e.g., \citealp{graham99,kelson97,kelson00}),
the Virgo Cluster is a venue where the interaction of galaxies with
the intracluster medium can be observed with relative ease.  
Recent X-ray observations have shown that the global underlying
emission arises from the hot intracluster gas 
(e.g., \citealp{es91,bohringer94,irwin96,schindler99}).
Spiral galaxies in Virgo are found to be \hi\ deficient relative to
spirals in the field, likely as a result of ram-pressure stripping
(e.g., \citealp{cayatte94,kk99,veilleux99,vollmer03}).

Due to their lower gravitational potentials, dwarf galaxies ought
to be more sensitive to their surroundings and should especially be
less able to retain their gaseous contents.
Evaluating the degree of gas deficiency with galaxy mass can help
constrain models of interactions between the intracluster medium and
the interstellar medium.
It had been claimed previously that Virgo dwarf galaxies were
not more gas-poor than Virgo spirals (e.g., \citealp{hoffman88}).
This was not consistent with ram-pressure stripping, which should
yield larger gas deficiencies for dwarf galaxies \citep{hg86}.
However, similar gas deficiencies for spiral and dwarf galaxies had
been claimed as being more consistent with turbulent viscous stripping
(e.g., \citealp{nulsen82,hg86}; see also \citealp{kenney90}).
For gas-rich dwarf galaxies in the Virgo Cluster, the condition for
ram-pressure stripping can be satisfied (e.g., \citealp{gh89,irwin96}).
The present focus is on examining dwarf irregular galaxies (dIs).
dIs are chosen over blue compact dwarf galaxies (BCDs), because the
light contribution from the underlying old stellar population can be 
separated from the contribution from young stars to obtain better
measures of the stellar mass. 
% A control sample of nearby (hereafter, ``field'') dIs was constructed
% and described in Paper~I \citep{lee03}.
 
Environmental effects on dwarf galaxies can be evaluated by examining:
(1) whether the interstellar gas from dwarfs is removed;
(2) whether the removal of neutral gas affects subsequent evolution;
and 
(3) whether these effects are manifested in diagnostic diagrams
such as the metallicity versus galaxy luminosity 
diagram (see, e.g., \citealp{rm95}) or the metallicity
versus gas fraction diagram (see, e.g., \citealp{llp01}).
There does in fact appear to be differences between dwarf elliptical
galaxies in clusters and similar nearby dwarf galaxies \citep{bb02}.
Answers to the questions above are the objectives of the present work, 
which focuses on Virgo Cluster dIs.

For a sample of dIs located in the core and the periphery of the 
Virgo Cluster, spectroscopic data are obtained with the following
goals: 
(1) where possible, to measure \othreea\ to derive direct oxygen
abundances; and 
(2) if \hii\ regions are too faint, to obtain spectra with sufficient
wavelength coverage to derive indirect abundances.
Derived oxygen abundances are combined with available data in the
literature to compare the properties of Virgo dIs with those
of the control dI sample.

The present paper is organized in the following manner.
The control sample and the Virgo Cluster sample of dIs are
described in \S~\ref{sec_samples}.
Observations and reductions are described in \S~\ref{sec_obs}.
The analysis of the spectra is presented in \S~\ref{sec_spectra}.
Nebular abundances are discussed in \S~\ref{sec_nebular} and
derived properties are presented in \S~\ref{sec_derivedprops}.
Discussion about the diagnostic diagrams and the evolution of
Virgo dwarfs are given in \S~\ref{sec_diagnostics} and
\S~\ref{sec_discussion}, respectively.
A summary is given in \S~\ref{sec_concl}. 

\section{The Samples of Dwarf Irregulars}
\label{sec_samples}

\subsection{The Field Sample}

Field dIs are dwarf galaxies with generally few neighbours.
Actually, field dIs are normally members of nearby loose groups
(e.g., IC~342/Maffei, Centaurus~A, Sculptor) and there are few truly
``isolated'' field dIs.
However, the evolution of field dwarfs is not likely to be complicated 
by external effects, such as those found in clusters of galaxies,
which are populated with thousands of galaxies and contain a
significant mass of gas in the intracluster medium.
So, field dwarf irregulars provide an excellent control
sample against which a sample of cluster dwarf galaxies can be
compared to evaluate environmental effects on galaxy evolution. 
The field sample of dIs was drawn from \cite{rm95} and updated with
recent distances and spectroscopic data obtained from the
literature.
A discussion of the sample was presented in Paper~I \citep{lee03}.

\subsection{The Virgo Cluster Sample}

For the most part, there is a lack of high-quality spectra 
with sufficient wavelength coverage for dIs in the Virgo Cluster,
although some data have appeared in the literature
(\citealp{kd81,hi86,sl88,slc88,gh89,ig90,vilchez95,gavazzi02,pustilnik02})\footnote{
See also Appendix~\ref{sec_appb}.}.
However, these samples studied suffer from the following:
``pure'' BCDs were included;
\othreea\ was not measured in any of the dIs; 
the spectral resolution was too low;
and 
the spectral coverage was incomplete, which meant that not even an 
abundance using other methods (\S~\ref{sec_brightline}) could be
derived.

A sample of dIs was constructed using the Virgo Cluster Catalog 
(VCC; \citealp{vcc,bts87,bpt93}). 
Targets were selected according to the following criteria:
(1) they must have been clearly identified as members of the 
cluster \citep{vcc,bc93,bpt93};
(2) they had to contain suitably bright \hii\ regions, as shown 
by published \halpha\ images \citep{gh89,ab98,hab99};
(3) they were located in the central and outer regions of the
cluster to sample high-density and low-density conditions, respectively;
(4) they lay within a narrow luminosity range 
($-15 \ga M_B \ga -17$), so that even a small offset from the
metallicity-luminosity relation for field dwarfs could be flagged
statistically;
and
(5) there was a range of over $\sim 100$ in \hi\ mass at a
given luminosity. 

Figure~\ref{fig_vdi} shows where each dI in the Virgo sample is
located in the cluster.
Basic properties for the galaxies are listed in
Table~\ref{table_vccdi}.
All distances are on a scale where the distance modulus for the
Large Magellanic Cloud (LMC) is 18.58 \citep{panagia99}.
The adopted value for the distance modulus to the Virgo Cluster
is 31.12~mag, which is $+0.08$ mag higher than the \cite{ferrarese96}
value to account for the updated distance modulus to the LMC. 

\section{Observations and Reductions}
\label{sec_obs}

\subsection{Observations}

Optical spectra of \hii\ regions in a sample of 11 dIs were obtained
at the Kitt Peak National Observatory (KPNO) and 
the Canada-France-Hawaii Telescope (CFHT). 
The characteristics of the instrumentation employed at each telescope
are listed in Table~\ref{table_obsprops}. 
A log of the observations is given in Table~\ref{table_obslog}.

Three nights (1997 March 1--3 UT) were awarded at KPNO.
The first night was completely unusable due to inclement weather.
Data were successfully obtained on the subsequent two nights
and spectra were obtained for seven dIs.
Moderate-resolution optical spectroscopy was obtained using the
Ritchey-Chr\'etien Spectrograph on the 4.0-meter telescope
in the f/7.8 configuration.
The spectrograph was used in long-slit mode for two-dimensional
spectroscopy to cover the maximum possible number of \hii\ regions.
The extreme ends of the CCD could not be used due to vignetting and
defocusing effects. 
Figures~\ref{fig_gh_v0848} to \ref{fig_gh_v1585v1789} show the
placement of the long slit over each dI observed at KPNO.
Individual exposures were 30~minutes and total exposure times
ranged between one and two hours.

A single 1800-second ``dark'' frame was obtained to
evaluate the contribution from the dark current.
An inspection of the dark frame showed that the total counts
registered were negligible compared to the counts from the sky
background in an image of comparable exposure time.
Thus, no corrections for dark current were applied.

To correct for variations in the pixel-to-pixel sensitivity of the
CCD, flat-field exposures of a quartz lamp within the 
spectrograph were taken at the beginning and end of each night.
Twilight flats were acquired at dusk each night to correct
for variations over larger spatial scales.
To correct for the ``slit function'' in the spatial direction, the
variation of illumination along the slit was taken into account
using internal and twilight flats. 

Two exposures of a helium-neon-argon (He-Ne-Ar) arc lamp were
acquired on each night for wavelength calibration. 
Flux calibration was achieved by observing the standard stars 
G191B2B and Feige~67 (see \citealp{oke90}) interspersed among 
observations of dIs. 
For all standard star exposures, the position of the star was set 
near the center of the slit. 

Three nights (1999 April 9--11 UT) were awarded at CFHT.
Spectra for two dIs were obtained on the first night until 
thick clouds prevented further observing.
The dome was not opened on the second night due to fog, snow, and
ice. 
Spectra for another three galaxies were obtained on the final night
until about local midnight, when fog and high humidity 
ended the observing run. 

Optical spectroscopy was obtained using the Multi-Object Spectrograph
(MOS) in the f/8 configuration.
The MOS spectrograph uses focal-plane masks created from previously 
obtained images \citep{lefevre94}. 
For each galaxy, five-minute \halpha\ exposures were obtained in
imaging mode to locate possible \hii\ regions.
Figures~\ref{fig_v0512v1114cfh} to \ref{fig_v1249v1554cfh} inclusive
show raw five-minute images in \halpha\ of four galaxies for which
spectra were successfully obtained.
An \halpha\ exposure for VCC~1448 did not reveal any compact
\hii\ regions at the position of the galaxy, confirming the
observations of \cite{gh89}.

For each \halpha\ image, every potential \hii\ region was marked and
the coordinates were recorded into a text file. 
Slit masks were subsequently constructed on-line using a YAG laser.
Slits for each \hii\ region were constructed to be approximately
2\arcsec\ wide and 10\arcsec\ long. 
Sky slits of the same size were placed near target slits.
For pointing, holes with diameters 3\arcsec\ and 5\arcsec\ were
created in the mask to transmit faint and bright stars,
respectively, in the field of view.

Spectra of a neon-argon-mercury (Ne-Ar-Hg) arc-lamp
and the standard star Feige~67 were acquired to wavelength- and
flux-calibrate, respectively, the galaxy spectra.
Illuminated dome flat-field exposures were obtained to remove
pixel-to-pixel sensitivity variations. 
From the dome flats, a two per cent variation in raw counts was found 
from the edge to the center of the slit. 
Since five slit masks (one for each galaxy) were inserted into the
mask slide on a given night, there was insufficient time to acquire
twilight flats for all masks.

Spectra at the center of the field of view were aligned perpendicular
to the slit.
Spectra of \hii\ regions farther away from the center exhibited
pin-cushion distortion.
The spectral coverage varied for each slit placement, depending on 
its location within the field of view.
For the most part, the spectral coverage was between 3600~\AA\
and 7200~\AA.
For VCC~1249 (UGC~7636), the position of the \hii\ region, 
LR1 (see \citealp{lrm00}), was near the top of the field
(Figure~\ref{fig_v1249v1554cfh}).
Because of the non-central position of the slit used for LR1, the
spectral range was cut off redwards of 6700~\AA.

\subsection{Reductions}

Long-slit spectra were reduced in the standard manner using 
IRAF\footnote{
IRAF is distributed by the National Optical Astronomical
Observatories, which is operated by the Associated Universities for
Research in Astronomy, Inc., under contract to the National Science
Foundation.
} 
routines.
Twilight flats and internal quartz lamp exposures were used to remove
the large-scale and the pixel-to-pixel variations in response,
respectively. 
Cosmic rays were identified and deleted manually.
Final one-dimensional flux-calibrated spectra for each \hii\ region
were obtained via unweighted summed extractions. 

Although reductions of the multi-slit data proceeded similarly as
those for the long-slit data, there were key differences.
Because of the small slit sizes with MOS, the slit function or
illumination across each slit was assumed to be constant over the
entire slit. 
Corrections for geometric distortions were not required for 
target spectra at the center of the frame.
A different task (e.g., {\em apflatten}) was used to remove
large-scale response variations.
For a given galaxy, subsections of dome flat exposures
corresponding to the positions of \hii\ regions were extracted.
For a given night, the sensitivity function was assumed to be constant
from one slit to the next slit.
An average response curve was obtained and used to flux-calibrate
the resulting \hii\ region spectra. 
Apertures corresponding to \hii\ regions were carefully defined 
to optimize signal quality.
Good background subtraction was difficult to achieve, because the slit
size was small and the background signal was dominated by the galaxy. 
Data were extracted for each aperture and averaged to yield
one-dimensional spectra; sky subtraction was performed
during extraction.
These spectra were subsequently corrected onto a linear wavelength
scale, and flux-calibrated using the standard star spectra.

Spectra observed for each Virgo dI are illustrated. 
The \othreea\ line was detected in \hii\ regions
VCC~0848-1 and VCC~1554-1.
These two spectra are shown in their entirety in 
Figure~\ref{fig_vccspca}.
For the remaining dIs where \othreea\ was not detected,
the spectra are shown in Figure~\ref{fig_vccspcb}.
The best spectrum for each dI is shown.
The following emission lines were observed in all spectra: 
\otwo, \hbeta, \othree, \halpha, \ntwob, and \stwo\ (except
VCC~1249-1 where the spectral range did not extend redwards
of 6700~\AA).

\section{Measurements and Analysis}
\label{sec_spectra}

Emission-line strengths were measured using locally developed
software. 
Because \othreea\ was detected in only two dIs, an upper limit to the
flux was estimated for the remaining dwarfs. 
A two-sigma upper limit to the \othreea\ flux was computed as the
product $2.51 A w$, where $A$ is two times the root-mean-square of
the underlying counts in the continuum immediately surrounding the
\othreea\ line, and $w$ is the full width at half maximum 
obtained from a fit of the nearest strong line.
Because of its proximity, the width of \hgamma\ was adopted
for the width of \othreea.

\subsection{Underlying Balmer Absorption}
\label{sec_balmerabs}

\cite{mrs85} found that stellar flux contributes most to the
continuum in the optical in the spectra of extragalactic \hii\ regions. 
Underlying Balmer absorption results in smaller Balmer emission line
fluxes, underestimates of Balmer emission equivalent widths, 
and overestimates of the reddening. 
Because emission lines are referenced to \hbeta, fluxes of forbidden
lines relative to \hbeta\ are overestimated if the underlying
Balmer absorption is not taken into account.
Most workers have reported in the literature a correction of 2~\AA\
for the underlying Balmer absorption (e.g., \citealp{mrs85}).

With sufficient spectral resolution and strong continua, a few
of the spectra exhibited obvious underlying Balmer absorption,
especially at \hbeta. 
It was possible to fit simultaneously at \hbeta\ the emission line and
the full absorption line.
Fits of emission and absorption profiles at \hbeta\ allowed for a
direct determination of the equivalent width of the underlying Balmer
absorption.
A measurement of the \hbeta\ equivalent width is applicable to any
Balmer line, owing to the flatness of the Balmer decrement
in absorption.

``Emission-only'' and ``emission-plus-absorption'' profiles were
fitted separately at \hbeta. 
First, a fit was constructed by assuming that the line consists only
of emission.
Second, an emission line profile and an absorption line profile were
simultaneously constructed.
All profiles were assumed to be integrated Gaussians\footnote{
An integrated Gaussian profile is a Gaussian profile
which is discretely sampled by bins of width equal to the width of
each pixel.
}.
Profile fits for the \hii\ region VCC~1179-1 are illustrated in
Figure~\ref{fig_linefits}, where fits at \hbeta, \othreeb, and
\othreec\ are shown.

The effective equivalent width of the underlying absorption at \hbeta\ 
is obtained by taking the difference of the emission equivalent width
of the ``emission-only'' line fit and the emission equivalent width 
determined from the ``emission plus absorption'' fit.
The resulting effective equivalent widths for underlying Balmer
absorption at \hbeta\ are given in Table \ref{table_eqwidths}.
From the values listed in column~(5), the average equivalent width for
underlying absorption at \hbeta\ is 
\begin{equation}
\langle \, W_{\rm abs}(\hbeta) \, \rangle 
        = 1.59 \pm 0.56 \;\: \mbox{\AA}.
\label{eqn_avg_ewabs}
\end{equation}
The error listed is the standard deviation.
This result is consistent with those obtained by 
\cite{mrs85}, \cite{diaz88}, and \cite{gd99}.
For spectra without a measurement of the underlying Balmer absorption, 
Equation~(\ref{eqn_avg_ewabs}) was used to correct for the
underlying Balmer absorption.
Subsequent reanalyses by setting the equivalent width of underlying
Balmer absorption to 2~\AA\ did not significantly change the results.

Column~(4) in Table~\ref{table_eqwidths} lists the equivalent width
of the entire absorption line profile at \hbeta. 
Results lie in the range between 3 to 7~\AA.
These values are consistent with what is expected for OB associations
for instantaneous burst models with a stellar initial mass function of
Salpeter type between 1 to 80 \msun\ \citep{mrs85,jaschek_book}. 
The models imply that the \hii\ regions have ages between 4 and 40~Myr
\citep{gd99}. 

\subsection{Correcting Line Ratios}

After correcting for underlying Balmer absorption, observed
line ratios were corrected for reddening.
The method by which reddening corrections are applied
is described in \citet{lee03}.
Observed flux and corrected intensity ratios are listed
in Tables~\ref{table_allvirgo_data1} to \ref{table_allvirgo_data4}
inclusive.
Where appropriate, two-sigma upper limits to \othreea\ are given.
In a number of cases, the adopted reddening was zero.
Uncertainties for the observed flux ratios account for the
uncertainties in the fits to the line profiles, their surrounding
continua, and the relative uncertainty in the sensitivity function
listed in Table~\ref{table_obslog}.
Uncertainties for the observed ratios do not include the uncertainty
in the flux for the \hbeta\ reference line.
Uncertainties in the corrected intensity ratios account for
uncertainties in the flux at the specified wavelength $\lambda$,
and at the \hbeta\ reference line; flux uncertainties are conservative
maximum and minimum values based upon approximate $2\sigma$
uncertainties in fits to emission lines.
However, uncertainties in the adopted reddening values are not
included.

\section{Nebular Abundances}
\label{sec_nebular}

\subsection{Oxygen Abundances}

\subsubsection{Direct (\othreea) Method}

The direct (or standard) method of obtaining oxygen abundances 
from emission lines is applicable to any galaxy where \othreea\ is
detectable and for which the doubly ionized O$^{+2}$ ion is the
dominant form of oxygen \citep{osterbrock_book}. 
The method by which oxygen abundances are derived with the direct
method is also described in \citet{lee03}.
The \othreea\ emission line was detected in VCC~0848-1 and VCC~1554-1
and direct oxygen abundances were subsequently derived.

Occasionally, forbidden emission from neutral oxygen (O$^0$) is
observed in \oonea\ and \ooneb\ lines.
Strong [O~I] emission indicates the presence of shocks from supernova
remnants (see, e.g., \citealp{skillman85}). 
However, there is very little oxygen in the form of O$^0$ in typical
\hii\ regions (see \citealp{bpt81,vo87,stasinska90}).
\oonea\ and \ooneb\ lines were detected in the spectra for 
VCC~0848-1 and VCC~1554-1.
The $I(\oonea)/I(\hbeta)$ values here are below ten per cent,
which are in good agreement with the range found in photoionization
models of metal-poor \hii\ regions \citep{stasinska90}.
The implied contribution by neutral oxygen to the total oxygen
abundance is about 0.02~dex.
The contribution from neutral oxygen was not included in the reported
values of the oxygen abundance for VCC~0848-1 and VCC~1554-1. 

\subsubsection{The Bright-Line Method}
\label{sec_brightline}

In the absence of \othreea, the empirical or ``bright-line'' method
was used to compute oxygen abundances.
The method is so called because the oxygen abundance is given in terms
of the bright [O~II] and [O~III] lines. 
\cite{pagel79} suggested that the index
\begin{equation}
R_{23} = \frac{I({\otwo}) + I({\othree})}{I({\hbeta})}
\label{eqn_r23_def}
\end{equation}
could be used as an abundance indicator.
\cite{mrs85} verified that this index was the best choice for
evaluating oxygen abundances empirically.
For the dwarf galaxies in the present work, the \cite{mcgaugh91}
calibration was used to derive oxygen abundances.

\cite{mcgaugh91,mcgaugh94} produced a set of photoionization models
using $R_{23}$ and 
\begin{equation}
O_{32} = \frac{I({\othree})}{I({\otwo})}
\label{eqn_o32_def}
\end{equation}
to estimate the oxygen abundance. 
However, $R_{23}$ is not a monotonic function of the oxygen abundance.
At a given value of $R_{23}$, there are two possible choices of the
oxygen abundance as shown in 
the left panel of Figure~\ref{fig_brightline}.
Each filled circle represents an \hii\ region from dIs in the 
control sample with measured \othreea.
Model curves from \cite{mcgaugh91} are superposed.
A long-dashed line marks the approximate boundary below (above) which 
the lower (upper) branch occurs.
Despite the fact the oxygen abundances for \hii\ regions in
dIs from the control sample range from about one-tenth to about
one-half of the solar value\footnote{
The solar value of the oxygen abundance of 12$+$log(O/H) = 8.87
\citep{zsolar} is adopted for the present work.
However, recent results (e.g., \citealp{zsolarnew}) have shown that
the solar value may in fact be smaller by $\approx$~0.2~dex.
}, 
these \hii\ regions are clustered around the ``knee'' of the curves,
where ambiguity is greatest about the choice of the appropriate branch
in the absence of \othreea.

Fortunately, $I$(\ntwob)/$I$(\otwo), or the \ntwootwo\ intensity
ratio, can discriminate between the lower and upper branches
\citep{mrs85,mcgaugh91,mcgaugh94,vanzee98sp}.
A plot of the \ntwootwo\ intensity ratio versus $R_{23}$ is shown in
the right panel of Figure~\ref{fig_brightline}.
The strength of the \ntwob\ line is roughly proportional to the
nitrogen abundance and the \ntwootwo\ intensity ratio is relatively
insensitive to ionization.
\cite{mcgaugh94} has shown that in galaxies ranging from sub-solar
to solar metallicities, \ntwootwo\ can vary by one to two orders of
magnitude and that \ntwootwo\ is roughly below (above) 0.1 at 
low (high) oxygen abundance.
For field dIs, lower-branch abundances obtained using the bright-line
calibration with \ntwootwo\ as a discriminant are consistent with
direct \othreea\ determinations.
The \ntwootwo\ indicator is clearly adequate to distinguish between
the lower and upper branches of the calibration.
\hii\ regions for Virgo dIs lie in the same locus as \hii\ regions for
field dIs, implying that Virgo dIs have lower branch abundances.

Analytical equations may be expressed for the oxygen abundance in
terms of $x \equiv \log\,R_{23}$ and $y \equiv \log\,O_{32}$.
The expressions for lower-branch and upper-branch oxygen abundances
are
%
% \begin{eqnarray}
% 12 + \log\,({\rm O/H})_{\rm lower} & = & 
% 12 - 4.93 + 4.25x - 3.35\sin \,(x) - 0.26y - 0.12\sin \,(y), \\
% %
% 12 + \log\,({\rm O/H})_{\rm upper} & = & 
% 12 - 2.65 - 0.91x + 0.12y \, \sin \,(x),
% \end{eqnarray}
%
\begin{eqnarray}              % FOR EMULATEAPJ
\lefteqn{\log\,({\rm O/H})_{\rm lower} = } \nonumber \\
& & -4.93 + 4.25x - 3.35\sin \,(x) - 0.26y - 0.12\sin \,(y), \\
\lefteqn{\log\,({\rm O/H})_{\rm upper} = } \nonumber \\ 
& & -2.65 - 0.91x + 0.12y \, \sin \,(x),
\end{eqnarray}
respectively, where the argument of the trigonometric 
function is in radians (McGaugh 1997, his ``model'' calibration,
private communication).
\cite{kobulnicky99} list a slightly different set of equations 
from McGaugh (his ``semi-empirical'' calibration); however, abundances
obtained from both sets of equations are similar. 

In Figure~\ref{fig_oxydiff}, oxygen abundances obtained directly
from \othreea\ measurements for \hii\ regions in the samples of field
and Virgo dIs are plotted against oxygen abundances obtained from the
lower-branch as required by \ntwootwo\ using the equations above.
The solid line marks where direct abundances are equal to empirical
abundances.
While the general trend is indeed correct, the scatter of the data
points about the line of equality is consistent with the 0.2~dex
uncertainty stated by McGaugh in determining oxygen abundances from
the bright-line method.
This is slightly less than the 0.3~dex uncertainty stated by
\cite{skillman89} for abundances computed below 
12$+$log(O/H) $\sim$~8.2.

% In the absence of \othreea, measurements of \otwo\ and \othree\ lines
% can be combined with the formulae above to obtain a measure of the
% oxygen abundance, and the value of \ntwootwo\ discriminates between
% upper and lower branches.
%
With the agreement seen between direct and bright-line abundances, 
the bright-line calibration combined with the \ntwootwo\ discriminant
can now be applied with confidence to dwarfs lacking \othreea\ detections,
such as the majority of Virgo dIs studied here. 
Figures~\ref{fig_brightline} and \ref{fig_oxydiff} have shown that the
nebular diagnostics observed in the \hii\ regions of Virgo dIs are
comparable to those found in the \hii\ regions of dIs in the control
sample.
A precise absolute calibration is not required for a 
{\em comparison\/} between Virgo dwarfs and dwarfs in the control
sample, because oxygen abundances for both datasets have been computed
in a similar manner using the same form of the calibration.
The primary goal here has simply been to look for an offset in
chemical abundances between Virgo and field dwarfs, which would
reflect the effect of differences in evolution. 

\subsection{Abundance of Nitrogen Relative to Oxygen}

The nitrogen-to-oxygen ratio, N/O, was estimated
from $N$(N$^+$)/$N$(O$^+$), if the temperature, $T_e$, were known
(e.g., \citealp{garnett90}). 
In the absence of \othreea, temperatures were derived with the
aid of the bright-line calibration.
Bright-line oxygen abundances were derived; 
ionization parameters were also computed with the same
calibration (McGaugh 1997, private communication).
An estimate of the temperature was obtained from a plot of 
temperature versus oxygen abundance for various values of the
ionization parameter made by \citet[][Fig.~7]{mcgaugh91}.
With an estimate for the electron temperature, N/O was computed 
in the following manner: 
%
% \begin{equation}
% \frac{N({\rm N})}{N({\rm O})} =
% \frac{N({\rm N}^+)}{N({\rm O}^+)} = 
% \frac{I({\rm [N\;II]}\lambda\,6583)}{I({\rm [O\;II]}\lambda\,3727)}
% \cdot 
% \frac{j({\rm [O\;II]}\lambda\,3727;\;n_e,\,T_e)}
% {j({\rm [N\;II]}\lambda\,6583;\;n_e,\,T_e)},
% \label{eqn_nplus_oplus}
% \end{equation}
%
\begin{eqnarray}                      %%% use for emulateapj mode
\frac{N({\rm N})}{N({\rm O})} 
& = & \frac{N({\rm N}^+)}{N({\rm O}^+)} \nonumber \\
& = & 
\frac{I({\rm [N\;II]}\lambda\,6583)}{I({\rm [O\;II]}\lambda\,3727)}
\cdot 
\frac{j({\rm [O\;II]}\lambda\,3727;\;n_e,\,T_e)}
{j({\rm [N\;II]}\lambda\,6583;\;n_e,\,T_e)},
\label{eqn_nplus_oplus}
\end{eqnarray}
where $j$(\ntwob; $n_e$, $T_e$) and $j$(\otwo; $n_e$, $T_e$) are
the volume emissivities of \ntwob\ and \otwo, respectively.
Emissivities were computed using transition probabilities and
collision strengths from the literature 
\citep{pradhan76,mb93,lb94,wiese96}.

\section{Derived Properties}
\label{sec_derivedprops}

Derived properties are presented in Table~\ref{table_vccspecprops}.
%
% Meaningful upper limits to the \othreea\ flux and subsequent
% lower limits to the oxygen abundance were derived
% for the remaining galaxies.
%
The list of derived properties include \hbeta\ intensities corrected
for underlying Balmer absorption and reddening, derived and adopted
values of the reddening, observed \hbeta\ emission equivalent widths,
derived widths of the underlying absorption at \hbeta, electron
densities and temperatures, oxygen abundances derived from \othreea\ 
measurements and from the bright-line method, and
nitrogen-to-oxygen abundance ratios.
Errors in oxygen abundances were computed from the maximum and minimum
possible values, given uncertainties in the line intensities; 
however, errors for both reddening and temperature were not included.

Ratios of $\alpha$-element (e.g., Ne, Ar, S) abundances to
oxygen abundances are constant with oxygen abundance, which is
expected from standard stellar nucleosynthesis.
Relative $\alpha$-to-oxygen abundances are derived for 
the two \hii\ regions with \othreea\ detections, because their
spectra have the best quality: 
VCC~0848-1:
log(Ne$^{+2}$/O$^{+2}$) = $-0.60 \pm 0.06$,
log(Ar$^{+2}$/O$^{+2}$) = $-2.19 \pm 0.07$;
VCC~1554-1:
log(Ne$^{+2}$/O$^{+2}$) = $-0.59 \pm 0.06$,
log(Ar$^{+2}$/O$^{+2}$) = $-2.27 \pm 0.06$.
The relative neon- and argon-to-oxygen abundances agree
with those of low-metallicity \hii\ regions in other dwarf galaxies
compiled by \cite{vanzee97} and \cite{it99}.

\section{Diagnostics of Evolution}
\label{sec_diagnostics}

\subsection{Oxygen Abundance versus $B$ Luminosity}
\label{sec_z_mb}

For field dIs, metallicity as represented by oxygen abundance
is well correlated with luminosity.
From \cite{lee03}, the fit to the control sample is expressed by
\begin{equation}
12 + \log{\rm (O/H)} = (5.59 \pm 0.54) + (-0.153 \pm 0.025)\,M_B.
\label{eqn_z_mb_fit}
\end{equation}
This fit is shown as a solid line in Figure~\ref{fig_z_mb}.
Equation~(\ref{eqn_z_mb_fit}) is consistent with the relation
determined by \citet{rm95} for dwarfs brighter than $M_B = -15$.
Equation~(\ref{eqn_z_mb_fit}) will be adopted as the
metallicity-luminosity relation for all field dIs.
Oxygen abundances for the sample of Virgo dIs are consistent with
the control sample of dIs at comparable luminosities.
In particular, this diagram is unable to distinguish gas-normal
dwarf galaxies from gas-poor dwarf galaxies; see \S~\ref{sec_z_mu}.

\subsubsection{$B-V$ versus $B$ Luminosity}
\label{sec_bv_mb}

Photometric properties (i.e., luminosity and color) can reveal
whether Virgo dIs exhibit significant differences in star formation
rates compared to field dIs.  
Figure~\ref{fig_bv_mb} shows a plot of $B-V$ color versus absolute
magnitude in $B$ for the samples of field and Virgo dIs.
In disk-like systems, $B-V$ reddens by approximately $+$0.25~mag when
$M_B$ fades by 1~mag \citep{mb94}.
There is no significant color difference between field and Virgo dIs,
i.e., there has been no significant fading or brightening of Virgo
dwarfs with respect to field dwarfs, in agreement with the results
of \cite{gh89}.
$B-V$ colors do not vary with luminosity in a systematic way, which
suggests that if ages are invariant, a constant percentage of the
baryonic mass is involved in star formation. 

\cite{popescu02} recently found evidence from far-infrared
observations that late-type dwarf galaxies, in particular BCDs, may
contain more (cold) dust by up to an order of magnitude than
previously thought. 
BCDs are found to have the coldest dust temperatures and the highest
dust mass surface densities (normalized to the optical area), which
suggests that dust might form preferentially within
the more luminous starbursts seen in BCDs.
This lends further weight to studying dIs which have less extinction.
For the field and Virgo samples of dIs in Figure~\ref{fig_bv_mb},
there is no difference in dust content which is readily
discernible.
If the dust-to-gas ratios in \hii\ regions of Virgo dIs were
different than in field dIs, our computations of oxygen abundances
would have to be reconsidered.

\subsection{Oxygen Abundance versus Gas Fraction}
\label{sec_z_mu}

We consider the relative gas content, as the metallicity-luminosity
relationship does not allow for effects on gas content to be shown.
The fraction of baryons in gaseous form is a fundamental parameter,
because it determines the metallicity in models of 
chemical evolution (e.g., \citealp{pagel_book} and references
within).
The baryonic gas fraction is defined as the ratio of the gas mass
to the total baryonic mass in gas and stars; the gas fraction is 
written as $\mu = M_{\rm gas}/(M_{\rm gas} + M_{\ast})$.
The method by which stellar mass-to-light ratios, stellar masses, and
gas fractions were derived is described in \cite{lee03}. 
A comparison of our computed stellar mass-to-light ratios for
both samples of dIs with recent models for star-forming galaxies 
is discussed in Appendix~\ref{sec_appa}.

A relationship between oxygen abundance and the baryonic
gas fraction for the control sample of field dIs was
established in Paper~I \citep{lee03}.
The fit to the control sample was expressed by 
\begin{equation}
12 + \log ({\rm O/H}) = 
	(8.64 \pm 0.40) + (1.01 \pm 0.17)\,\log \,\log (1/\mu).
\label{eqn_z_mu_fit}
\end{equation}
Equation~(\ref{eqn_z_mu_fit}) is taken as the ``best fit'' 
and is shown as a solid line in Figure~\ref{fig_z_mu}. 
A key result from Paper~I was that the chemical evolution of
dIs in the control sample is consistent with little to no 
gas flows into or out of the dIs\footnote{
The reduced effective oxygen yield for the field dIs \citep{lee03},
however, may be more indicative of ``leaky boxes.''
}.

\subsubsection{Gas Deficiency Index for Dwarf Irregulars}
\label{sec_gdi}

As the cluster environment is expected to be detrimental to the
gas content in dwarf galaxies (as seen in the outer
regions of spirals in Virgo; e.g., \citealt{kk99}), one might
anticipate an offset between dIs in the Virgo Cluster sample and 
dIs in the control sample in a diagnostic which correctly gauges 
relative gas contents.
In Figure~\ref{fig_z_mu}, a few Virgo dIs appear in the
same locus of the metallicity-gas fraction diagram as field dIs,
whereas a number of Virgo dIs have noticeably lower gas fractions
compared to field dIs at a given oxygen abundance.  
In fact, through this diagram, it is now possible to construct an
index which quantifies the gas deficiency of a dI in order to compare
``gas-deficient'' dIs with isolated dIs. 

\citet[][also \protect \citealp{gh85}]{hg84} 
defined a deficiency parameter, DEF, for disk galaxies,
which compares the observed \hi\ mass of a galaxy, 
$M_{\rm HI}^{\rm obs}$, with the value expected for an isolated galaxy
of the same morphological type, $T^{\rm obs}$, and 
optical linear diameter $D_{\rm opt}^{\rm obs}$.
Isolation was defined on the basis of association with other 
galaxies \citep{kara73}, where a galaxy of angular diameter $d$ had to
lie at a projected distance of at least $20d$ from another galaxy.
The \citeauthor{hg84} deficiency parameter is written as
\begin{equation}
{\rm DEF} = 
\langle \log M_{\rm HI}(T^{\rm obs}, D_{\rm opt}^{\rm obs}) \rangle 
- \log M_{\rm HI}^{\rm obs}.
\label{eqn_def_defn}
\end{equation}
Using Equation~(\ref{eqn_def_defn}), \cite{hoffman87} showed that
Virgo dIs within about 5\degr\ of the cluster center were
deficient in \hi\ by roughly a factor of two compared to dIs outside
this circle, although the degree to which dIs were gas deficient 
was not more severe than for spirals \citep{hoffman88}.
Also, a number of workers have computed DEF for very late-type spirals
and irregulars using the relationship inferred from Sc spirals,
even though ``normal'' \hi\ content is well-defined only for spirals 
of type Sa to Sc (see, e.g., \citealp{solanes01}). 
The global oxygen deficiency parameter discussed by \cite{pf98}
applies to spiral galaxies and includes a dependency of their
prescribed gas fraction with galactocentric radius.
However, to determine and quantify evolutionary effects within dIs, 
gas-poor dwarfs should be compared with gas-normal dwarfs. 

Here, a gas-deficiency index (GDI) for dwarf irregulars is defined by 
\begin{equation}
{\rm GDI} \equiv 
	\log \left[ \frac{ \left( M_{\rm gas}/M_{\ast} \right)_p }
	{ \left( M_{\rm gas}/M_{\ast} \right) } \right],
\label{eqn_gdi_defn1}
\end{equation}
where the subscript ``$p$'' refers to the prediction for an isolated
dwarf. 
In terms of the gas fraction, GDI can be expressed as
\begin{equation}
{\rm GDI} = \log \left( 
	\frac{\mu_p}{\mu} \cdot \frac{1 - \mu}{1 - \mu_p}
	\right),
\label{eqn_gdi_defn2}
\end{equation}
where $\mu$ is the observed gas fraction, and $\mu_p$ is the gas
fraction predicted from the measured oxygen abundance using the 
fit expressed by Equation~(\ref{eqn_z_mu_fit}).
Deficiency increases with increasingly positive GDI.
The present form of the GDI has the following advantages:
(1) it is founded exclusively upon the properties of gas-rich dwarf
galaxies and there is no mixing of morphological types, 
e.g., Sc's with Sd's or with dI's; 
(2) the index is independent of absolutes, such as size; and
(3) the index is distance-independent.

Table~\ref{table_alldi_starsgas} lists stellar masses, gas masses, and
GDIs for the samples of field and Virgo dIs.
Although no \hii\ region spectrum was measured for VCC~1448,
it is included here to increase the number of gas-poor dIs in the 
sample; the oxygen abundance for VCC~1448 was estimated from the 
metallicity-luminosity relation (Equation~\ref{eqn_z_mb_fit}).
Histograms for the samples of field dIs and Virgo dIs are shown in
Figure~\ref{fig_gdi_hist}.
As expected, field dIs are mostly clustered around the value 
GDI $\simeq 0$. 
Seven Virgo dIs have GDIs in the range between $-0.5$ and $+0.5$,
and the other five Virgo dIs have GDI $\ga +0.8$.
A few Virgo dIs are gas-deficient by a factor of about 30 
(GDI $\simeq +1.5$); only three per cent of the expected gas content has
remained.
A Kolmogorov-Smirnov test on the two distributions of GDI
returned the statistic $D = 0.576$ with a significance level
$P = 6.3 \times 10^{-3}$ ($N = 22$ field dIs; $N = 12$ Virgo dIs).
Thus, the two data sets are significantly different and can not be 
drawn from the same population. 
For the present discussion, a ``gas-poor Virgo dI'' is defined 
to be one with GDI $\ga +0.8$.

It is instructive here to compare the present GDI with the previous
measure of gas deficiency (DEF; Equation~\ref{eqn_def_defn}). 
For example, \cite{kk01} computed a DEF $= -0.30$ for VCC~1554,
which is not very different from the present GDI value ($-0.38$).
The scales are consistent, at least for more luminous and massive
irregular galaxies.

\subsection{Nitrogen-to-Oxygen Abundance Ratios}

Based upon observations of \hii\ regions in spiral and dwarf
galaxies, nitrogen appears to be both a primary and secondary product
of nucleosynthesis. 
It remains uncertain, however, whether nitrogen is produced mostly
from short-lived massive stars or from longer-lived
intermediate-mass stars. 
The closed-box model \citep{ep78,pagel_book} predicts that the
nitrogen-to-oxygen abundance ratio, N/O, is constant with 
the oxygen abundance if nitrogen is of ``primary'' origin, and 
is proportional to the square of the oxygen abundance
if nitrogen is of ``secondary'' origin.

Measurements of the nitrogen-to-oxygen ratio have been used to
differentiate between the different origins for nitrogen (see, e.g.,
\citealp{garnett90}). 
It has been suggested that N/O can be used as a ``clock'' to
measure the time since the last burst of star formation 
(e.g., \citealp{ep78,garnett90,sbk97,henry00}).
This would work if bursts of star formation were separated by long 
quiescent periods and if the delivery of nitrogen into the
interstellar gas were delayed relative to oxygen.

A plot of log(N/O) versus log(O/H) is shown in Figure~\ref{fig_no_oxy}
for \hii\ regions in the sample of field dIs with published data and
and for \hii\ regions in the sample of Virgo dIs. 
The two sets of galaxies overlap, although some \hii\ regions in Virgo
dIs appear to have somewhat elevated values of log(N/O) for their oxygen
abundances.
A larger set of data for \hii\ regions in a mixture of other dIs
and BCDs \citep{garnett90,ks96,vanzee97,it99} is included.
Although the data are rather heterogeneous, they illustrate that the
range of log(N/O) values for the Virgo sample is not anomalous
compared to the field sample.

Figure~\ref{fig_no_oxy} shows that nitrogen in metal-poor dwarf
galaxies is likely of primary origin. 
The considerable scatter in N/O at a given O/H may be explained by the
time delay between the release of oxygen by massive stars and
nitrogen by intermediate-mass stars 
\citep{garnett90,sbk97,vanzee98nit,henry00}.
It is not clear that gas flows have a bearing on the scatter 
\citep{henry00}, because Figure~\ref{fig_z_mu}
(Equation~\ref{eqn_z_mu_fit}) showed that the chemical evolution of
field dIs is consistent with systems with negligible gas flows.
For the Virgo sample, there is little difference between \hii\ regions
in gas-normal dIs and \hii\ regions in gas-poor dIs.

The two \hii\ regions in Figure~\ref{fig_no_oxy} with unusually low
log(N/O) ($\la -1.9$) are UGC~5764-3 \citep{vanzee97} and VCC~1249-1
\citep{lrm00}. 
For UGC~5764, \cite{vanzee97} acknowledge that their \hii\ region \#3
may be unusual compared to the other two \hii\ regions in the galaxy.
Deeper follow-up spectroscopy of VCC~1249-1 would be desirable;
see \cite{lrm00}.

\section{Discussion} 
\label{sec_discussion}

\subsection{Effects of the Intracluster Medium}

As disk galaxies move at high speeds relative to the hot,
dense intracluster medium (ICM), ram-pressure stripping occurs when
the resulting pressure exceeds the self-gravity of the disk, and
removes the interstellar medium from galaxies \citep{gunngott72}.
Ram-pressure stripping by the intracluster medium in Virgo was
discussed by \cite{lrm00} to help explain the observations for
VCC~1249.
With the construction of a gas-deficiency index for dIs and a clear
indication that there are both gas-poor and gas-normal dIs in Virgo,
one can seek a correlation between the GDI and the density of
intracluster gas.

Figure~\ref{fig_vdi} showed the locations of the sample of dIs within
the Virgo Cluster. 
The left panel of Figure~\ref{fig_vdi_gasdef} illustrates the
positions of gas-deficient dIs (i.e., dIs with ${\rm GDI} \ga +0.8$)
with respect to the density of the ICM. 
Notably, these dIs happen to be found in regions
where (1) the distribution of gas in the ICM is higher than average
(as seen in X-ray surface brightness), and 
(2) the number of H~I-deficient spiral galaxies in the Virgo
Cluster is largest \citep{solanes01}.

To look for a correlation with the ICM density, X-ray surface
brightnesses at the positions of Virgo dIs were 
extracted from a map of the Virgo Cluster kindly provided by Sabine
Schindler (2001, private communication; \citealp{schindler99}).
The right panel of Figure~\ref{fig_vdi_gasdef} shows a plot of the
gas-deficiency index 
versus the X-ray surface brightness of the ICM.
There appears to be a threshold in X-ray surface brightness above
which gas-poor dwarfs lie.
Thus, the gas deficiency appears to be related to ICM gas density.

A correlation between the GDI and X-ray surface brightness 
would support the notion that stripping occurred recently.
\cite{gh89} claimed that most dIs have never passed through the 
cluster core, on the basis that a small fraction of present-day
dIs currently exhibit any signs of gas loss and that their $B-V$
colors are not significantly redder than those of field dIs, as
the latter is confirmed by Figure~\ref{fig_bv_mb}.
If stripping had occurred long ago, one would expect to find faint and
red gas-poor dwarfs occupying the upper left portion of
Figure~\ref{fig_bv_mb}; i.e., offset with respect to field dIs.
Given that $B-V$ should redden by about 0.4~mag in about 1~Gyr
\citep{bothun82,mb94}, the stripping event for the gas-poor
dwarfs must have occurred at most 1~Gyr ago. 
Moreover, gas-poor dIs should have already made a number of passages
through the core of the Virgo Cluster.
Using the formula in \cite{cote97} and the list of Virgo Cluster
member galaxies in \cite{bpt93}, the time for a typical galaxy to
cross the cluster is about 1~Gyr.  

Dwarf irregulars may be passing through the dense regions of the
ICM on orbits with large radial components
\citep{dressler86,giraud86}, although \cite{vandermarel00} find
that, on average, galaxy orbits within clusters might be isotropic.
On the other hand, dIs are unlikely to be in circular orbits in the
cluster for two reasons.
First, if their orbital radii were large, they would never encounter
the dense ICM gas, and there would be no gas-deficient dIs,
which is contrary to the observations.
Second, if their orbital radii were small, the dIs would always
encounter the dense ICM gas, and they would have been stripped 
long ago. 

The present level of chemical enrichment of gas-poor dIs in Virgo
likely was achieved well before the dIs were stripped, as the mass of
the underlying population of old stars is considerable.
On average, the mass in old stars for a dI from the present samples is
approximately $10^9 \, \msun$ (see \citealp{lee01}).
Based upon typical star formation rates for gas-rich dIs in Virgo
(e.g., \citealp{hab99}), and assuming that most of the stars 
formed at a rate which has been constant on average
(e.g., \citealp{vanzee01}), the time required to form the old
population is at least 10~Gyr. 
Thus, the bulk of the stars within Virgo dIs formed well in the past,
when the galaxies were in the outer regions of the cluster.
All of this suggests that dIs are falling into the central
regions and encountering the dense ICM gas for the first time;
see also \cite{conselice01}.

\subsection{Fading of Gas-Poor Dwarf Irregulars}

Since gas-poor dIs are recognizable, it is logical to ask what they
will become after fading, and whether examples of such galaxies exist
in the present-day. 
In attempting to ascertain what the dIs will become, comparisons are
made with other gas-poor dwarf galaxies in the Virgo Cluster, i.e.,
non-nucleated dwarf ellipticals (dEs), nucleated dwarf ellipticals
(dE,Ns), ``transition'' dwarfs  (dI/dE), and dIs with \hi\ fluxes
comparable to or smaller than that of VCC~1249 (gas-poor dwarfs;
\citealp{hoffman87,bc93}).

Fading vectors for gas-poor Virgo dIs are shown in
Figure~\ref{fig_fading}, where the exponential scale length and
effective surface brightness are both plotted against $M_B$.
In Figure~\ref{fig_fading}a, the scale lengths for gas-poor Virgo dIs
appear generally similar to other dIs discussed by \cite{pt96}.
They are indistinguishable from those of dEs, dE,Ns, ``transition''
dwarfs, and other gas-poor dwarfs at comparable luminosities.
The scale lengths for VCC~1249 and VCC~1448 are about a factor of two
larger than for the other gas-poor dIs. 
Indeed, VCC~1249 and VCC~1448 are situated very close in projection to 
the supergiant elliptical galaxies M~49 and M~87, respectively.
Tidal effects from the neighbouring ellipticals have likely caused 
the stellar components of VCC~1249 and VCC~1448 to expand.
In Figure~\ref{fig_fading}b, five dIs with ${\rm GDI} \ga +0.8$ lie on
the upper envelope of the locus defined by dEs/dSphs and other
gas-poor dwarfs identified solely by their low \hi\ content 
\citep{hoffman87}.
These five gas-poor dIs appear to be structurally similar 
to dEs and dE,Ns. 
However, dIs do not have prominent nuclei; it is not clear how
non-nucleated dIs can form compact nuclei.
The progenitors of nucleated Virgo dEs were not likely to have been
dIs.

A typical gas-poor dI in Virgo could fade by about 0.6 to 0.7~mag,
which would be expected to occur in about 1~Gyr \citep{bothun82}.
The lowest luminosity dEs ($M_B \approx -12$, $\mu_B \approx 26$) 
are about 3 to 4~mag fainter than the gas-poor dIs with
GDI $\ga +0.8$.
Assuming that the fading rate is constant and that the progenitors of
present-day dEs were similar to the gas-poor dIs seen today, the
gas-removal event for present-day dEs would have occurred $\approx$
5--6 Gyr ago.

Gas-poor Virgo dIs will evolve to become galaxies resembling dEs/dSphs
with scale lengths comparable to present-day dEs/dSphs, assuming
dwarfs are dark-matter dominated.
In all likelihood, there should be successors to stripped dIs 
in the cluster, and/or present-day progenitors to cluster dEs.
However, using spatial and velocity distributions of Virgo dwarf
galaxies, \cite{conselice01} argue that dEs presently seen in
the core cannot originate from accreted dwarf galaxies from the field,
but instead, evolved from some precursor population of galaxies with 
different morphology which fell into the cluster some time in the
past.
If this were strictly true, there would be very few post-stripped dIs
seen in the cluster; otherwise, the alternative for which some
evidence has been presented above is that some of the dEs in Virgo
must be stripped dIs that have subsequently faded.
Additional information about stellar populations of other cluster dEs
(e.g., \citealp{conselice03}) shows that the stripping of dIs remains
very much a viable option, and will help to constrain better the
processes responsible for the production of cluster dEs.

\section{Conclusions} 
\label{sec_concl}		% CONCLUSIONS

A sample of dIs in the Virgo Cluster was constructed to examine
the impact of the cluster environment on galaxy evolution.
Optical spectroscopic data were obtained for eleven Virgo dIs at KPNO
and CFHT.
\othreea\ was detected in VCC~0848 and VCC~1554 for which oxygen
abundances could be directly determined. 
Oxygen abundances for the remaining dIs were derived using the
bright-line method.
The line ratio $I$(\ntwob)/$I$(\otwo) was used as a discriminant to
break the degeneracy intrinsic to the method. 
Comparing with direct determinations in a sample of dIs in the field,
oxygen abundances obtained from the bright-line method were found to
be accurate to within $\approx$ 0.2~dex, with no obvious systematic
differences. 

Oxygen abundances for the sample of Virgo dIs were comparable to those
for field dIs at similar luminosities.
In the metallicity-gas fraction diagram, gas fractions for a number of
Virgo dIs were comparable to those for field dIs at similar oxygen
abundances, while a few Virgo dIs were noticeably gas-poor.
A new gas-deficiency index (GDI) for dwarfs was defined using the
observed relation between oxygen abundance and gas fraction for field
dIs.
A comparison of indices for field and Virgo dwarfs directly led
to the quantitative identification of five gas-poor Virgo dIs
with GDI $\ga +0.8$; a few of the latter are gas deficient by a factor
of 30.
A Kolmogorov-Smirnov test showed that the field and Virgo samples 
could not have been drawn from the same distribution.
The gas deficiency was found to be correlated with X-ray surface
brightness, which suggests ram-pressure stripping as the most
likely cause for gas removal.
Together with the observed colors and the lack of significant fading,
the results suggest that dIs are traversing the dense regions of the
cluster for the first time and are being stripped of their gaseous
content without significant effect on their luminosities and
metallicities. 
Subsequent gas-poor dIs will likely fade into systems resembling
present-day dwarf ellipticals in the cluster core.

\acknowledgments		% ACKNOWLEDGMENTS

The work presented here was part of a Ph.D. thesis completed
at York University in Toronto, Canada.
The authors thank the referee, Jeff Kenney, for constructive 
comments which greatly improved the presentation of this paper.
H. L. thanks Marshall McCall for support, and acknowledges the
Faculty of Graduate Studies and the Graduate Student Association at
York University for travel funds.
H. L. acknowledges Trinh Thuan, Noah Brosch, Giuseppe Gavazzi, 
Gerhard Hensler, Alexis Finoguenov, Jay Gallagher, Evan Skillman, 
Eric Bell, and Hans B\"ohringer for various insights arising
from discussions.
H. L. also thanks Deidre Hunter for a copy of \halpha\ images, 
Stacy McGaugh for numerical grids in computing oxygen abundances,
Bruno Binggeli for an electronic copy of the Virgo Cluster
Catalog,
Stephane Charlot for help in the use of their population synthesis
models, and 
Sabine Schindler for ROSAT maps of the Virgo Cluster.
M. L. M. would like to thank the Natural Sciences and Engineering
Research Council of Canada for its continuing support. 
M. G. R. gratefully acknowledges travel support from Marshall McCall,
from the Instituto de Astronom\'{\i}a at UNAM, and from 
CONACyT project 37214-E.
The authors wish to thank the staff at Kitt Peak National Observatory
and the Canada-France-Hawaii Telescope for their assistance with the
observations.
Some data were accessed as Guest User, Canadian Astronomy Data Center,
which is operated by the Dominion Astrophysical Observatory for the
Research Council of Canada's Herzberg Institute of Astrophysics.
This research has made use of the NASA/IPAC Extragalactic Database
(NED) which is operated by the Jet Propulsion Laboratory,
California Institute of Technology, under contract with the
National Aeronautics and Space Administration. 

\begin{appendix}
% \appendix	                            % APPENDIX

\section{Stellar mass-to-light ratios in B versus B-V}
\label{sec_appa}

For models of star-forming disk galaxies with reasonable 
metallicities and star formation histories, \cite{belldejong01} showed
that the stellar mass-to-light ratio correlated strongly with
the optical colors of the integrated stellar populations.
In Paper~I \citep{lee03}, a plot of the stellar mass-to-light
ratio in $B$ versus the absolute magnitude in $B$ for the sample of
field dIs showed that our two-component method of computing stellar
masses gave results in agreement with Bruzual \& Charlot population
syntheses for model galaxies with a constant rate of star formation
and ages between 10 and 20 Gyr old.
For completeness, Figure~\ref{fig_mstarlb} shows a plot of the stellar
mass-to-light ratio in $B$ against $B-V$ for both field and Virgo
samples of dIs.
The reddest galaxy in the top-right corner is the 
Virgo gas-poor dI VCC~1448. 
For the range in color, there is little difference between 
the two samples. 
With additional models shown to facilitate further comparisons.
our two-component method gives roughly similar stellar mass-to-light
ratios in $B$ as those obtained from population synthesis models to
within about 10\%.

\section{Latest measurements}
\label{sec_appb}

% VCC 241 is listed as a dI: in NED, but Binggeli calls it an Sd
With their \othreea\ measurement, \cite{zasov00} derived for VCC~241
(classified dI or Sd) an oxygen abundance of $12+\log{\rm (O/H)} = 7.58$.
This is at present the most metal-poor dwarf galaxy known in the Virgo
Cluster.
\cite{vi03} have recently obtained measurements of 22 
star-forming dwarf galaxies in Virgo.
Their sample consists of a mixture of blue compact dwarf
galaxies, Magellanic spirals, amorphous dwarfs, and dwarf irregulars.
Their oxygen abundances for VCC~0848, VCC~1179, and VCC~2037 agree
with the abundances presented here.

\end{appendix}

\clearpage	% FIGURE CAPTIONS : \clearpage every 7 figcaptions

\begin{figure} % FIGURE 1
% \epsscale{0.5}
% \plotone{fig01.ps}
\caption{
Location of dIs in the Virgo Cluster sample.
Declination (vertical axis) versus Right Ascension (horizontal axis).
North is at the top and East is to the left.
Symbol sizes are not to scale.
Large open circles show the positions of five giant elliptical
galaxies in the cluster.
The crosses mark the locations of 12 dIs in the cluster sample;
additional labels indicate their numeric designations in the Virgo
Cluster Catalog (VCC; \citealp{vcc}).
}
\label{fig_vdi}
\end{figure}

\begin{figure} % FIGURE 2
% \epsscale{0.7}
% \plotone{Lee.fig02.ps}
\caption{
\hii\ regions in VCC 0848 (A1223$+$06):
$R$ and continuum-subtracted \halpha\ images
from \citet{gh89}. 
The field of view in each frame is approximately 87\arcsec\ by
87\arcsec.
North is at the top and East is to the right.
The orientation and placement of the long slit are indicated 
in the \halpha\ image.
}
\label{fig_gh_v0848}
\end{figure}

\begin{figure} % FIGURE 3
% \epsscale{1.}
% \plottwo{Lee.fig03a.ps}{Lee.fig03b.ps}
\caption{
\hii\ regions in VCC~0888 (8\degr26; left) and 
VCC~2037 (10\degr71; right): $R$ and continuum-subtracted \halpha\
images from \citet{gh89}.
Image size and orientation are the same as Figure~\ref{fig_gh_v0848}.
}
\label{fig_gh_v0888v2037}
\end{figure}

\begin{figure} % FIGURE 4
% \epsscale{1.15} 
% \plottwo{Lee.fig04a.ps}{Lee.fig04b.ps}
\caption{
\hii\ regions in 
VCC~1179 (IC~3412; top) and 
VCC~1200 (IC~3416; bottom): 
% VCC~1179 (IC~3412; left) and 
% VCC~1200 (IC~3416; right): 
$R$ and continuum-subtracted \halpha\ images from \citet{gh89}. 
Image size and orientation are the same as Figure~\ref{fig_gh_v0848}.
}
\label{fig_gh_v1179v1200}
\end{figure}

\begin{figure} % FIGURE 5
% \epsscale{1.15}
% \plottwo{Lee.fig05a.ps}{Lee.fig05b.ps}
\caption{
\hii\ regions in 
VCC~1585 (DDO~136; top) and 
VCC~1789 (A1236$+$05; bottom): 
% VCC~1585 (DDO~136; left) and 
% VCC~1789 (A1236$+$05; right): 
$R$ and continuum-subtracted \halpha\ images from \citet{gh89}.  
Image size and orientation are the same as Figure~\ref{fig_gh_v0848}.
}
\label{fig_gh_v1585v1789}
\end{figure}

\begin{figure} % FIGURE 6
% \epsscale{1.}
% \plottwo{Lee.fig06a.ps}{Lee.fig06b.ps}
\caption{
Five-minute \halpha\ images of VCC~0512 (left)
and VCC~1114 (right) with MOS at CFHT. 
Images have not been corrected for continuum emission.
North is at the top and East is to the left. 
The 5\farcm{1} by 5\farcm{7} frame is displayed as a ``negative'' so
black objects on the image indicate bright sources.
The white vertical stripe is due to a set of bad columns on the CCD.
Left panel: 
Two \hii\ regions (labelled 3 and 4) in VCC~0512 (12\degr25) are
identified, over which small east-west slits were placed for
spectrophotometry.
This image may be compared with $R$-band and continuum-subtracted
\halpha\ images for 12\degr25 obtained by 
\citet[Fig. 1, Plate 26]{gh89}.
Right panel:
Two \hii\ regions (labelled 1 and 2) in VCC~1114 (8\degr30) are
identified, over which small east-west slits were placed for
spectrophotometry. 
The white vertical stripe is due to a set of bad columns on the CCD.
This image may be compared with $R$-band and continuum-subtracted
\halpha\ images for 8\degr30 obtained by 
\citet[Fig. 1, Plate 21]{gh89}.
}
\label{fig_v0512v1114cfh}
\end{figure}

\begin{figure} % FIGURE 7
% \epsscale{1.}
% \plottwo{Lee.fig07a.ps}{Lee.fig07b.ps}
\caption{
Five-minute \halpha\ images with VCC~1249 (left) and VCC~1554 (right)
with MOS at CFHT.  
General comments about each image (e.g., size, orientation) are the
same as Figure~\ref{fig_v0512v1114cfh}.
Left panel: 
At the position of VCC~1249 (centre of the image), there is only
faint, diffuse \halpha\ emission.
To the northwest, a new \hii\ region (LR1; \citealt{lrm00})
was identified, over which a small slit with east-west alignment was
placed for spectrophotometry.
Right panel:
In VCC~1554, five large \hii\ regions (labelled 1 to 5) and two
additional emission-line objects (labelled 6 and 7) are identified,
over which small slits with east-west alignment were placed for
spectrophotometry. 
Slits for \hii\ regions 1--5 are drawn here for visibility.
% The position of \hii\ region ``1'' relative to the main body of the
% galaxy appears to be analogous to that of the 30~Doradus nebula in the
% Large Magellanic Cloud.
}
\label{fig_v1249v1554cfh}
\end{figure}

\clearpage

\begin{figure} % FIGURE 8
% \epsscale{0.9}
% \plotone{fig08.ps}
\caption{
Emission-line spectra of VCC 0848-1 (a and b) 
and VCC 1554-1 (c and d):
the observed flux per unit wavelength is plotted versus wavelength. 
In panels a and c, key emission lines are labelled; whereas in
panels b and d, spectra are magnified to highlight the 
\othreea\ emission line.
}
\label{fig_vccspca}
\end{figure}

\begin{figure} % FIGURE 9
% \epsscale{0.9}
% \plotone{fig09.ps}
\caption{
Same as Figure~\ref{fig_vccspca}, but for selected \hii\ regions in
other dwarf galaxies from the Virgo Cluster sample.
}
\label{fig_vccspcb}
\end{figure}

\begin{figure} % FIGURE 10
% \epsscale{1.}
% \plottwo{fig10a.ps}{fig10b.ps}
\caption{
Line profile fits at \hbeta, \othreeb, and \othreec\ for 
\hii\ region VCC~1179-1: observed flux per unit wavelength
(ergs~s$^{-1}$~cm$^{-2}$~\AA$^{-1}$) versus wavelength (\AA).
The dotted line connects consecutive data points.
The solid line is a fit to the lines and continuum.
Left panel: 
A profile fit at \hbeta\ is obtained assuming that the line
consists only of emission.  
The equivalent width of the emission line profile is 
$1.76 \pm 0.25$~\AA.  
Right panel: 
An emission line profile and an absorption line profile are
simultaneously fit at \hbeta.  
The equivalent widths of the emission line and the absorption line are
$3.71\,\pm\,0.23$~\AA\ and $5.25\,\pm\,0.48$~\AA, respectively.
Thus, the effective equivalent width of the underlying absorption at
\hbeta\ is $1.95\,\pm\,0.34$~\AA, which is the difference between the
emission equivalent width derived from the ``emission-only'' fit and
the emission equivalent width from the ``emission-plus-absorption''
fit. 
}
\label{fig_linefits}
\end{figure}

\begin{figure} % FIGURE 11
% \epsscale{1.1}
% \plottwo{fig11a.ps}{fig11b.ps}
\caption{
Bright-line method of deriving oxygen abundances.
Left panel:
Oxygen abundance versus $R_{23}$ for \hii\ regions in 
field and Virgo dIs.
Filled circles indicate for field dIs \hii\ regions with
direct (\othreea) oxygen abundances.
Direct abundances for VCC~0848-1 and VCC~1554-1 are shown as  
crosses.
The horizontal dotted lines indicate solar (12$+$log(O/H)$_{\odot}$ =
8.87; \citealt{zsolar}) 
and one-tenth solar values of the oxygen abundance.  
Calibration curves for log(O/H) against log~$R_{23}$
(McGaugh 1997, private communication) are plotted
for four different values of the ionization parameter.
These curves were derived using a standard stellar initial mass
function for a cluster of ionizing stars with an upper mass limit of 
${\cal M}_{\rm up}$ = 60~\msun (solid lines) and 
100~\msun (short-dash lines).
The horizontal long-dash line at log(O/H) $\simeq -3.6$
marks the boundary below (above) which the lower (upper) branch
approximately occurs.
\othreea\ abundances are consistent with lower-branch values.
The error bar at the lower right indicates typical uncertainties of
0.1~dex for the $R_{23}$ indicator and 0.1~dex for \othreea\
oxygen abundances.
Right panel:
log $[I(\ntwob)/I(\otwo)]$ versus log $R_{23}$ for \hii\ regions in 
field and Virgo dIs.
Symbols representing \hii\ regions for field and Virgo dIs are
the same.
The dotted lines mark roughly the regions occupied by high surface
brightness (HSB) spiral galaxies \citep{mrs85} at the upper left 
and low surface brightness (LSB) galaxies \citep{mcgaugh94} towards
the lower right.
The horizontal dashed line marks the boundary below (above) which 
the lower (upper) branch of the bright-line calibration is selected
to determine a unique value of an oxygen abundance.
%
% There are only three \hii\ regions whose \ntwootwo\ values are above the
% boundary and which imply bright-line abundances inconsistent with direct
% determinations.
% However, the dIs in which these \hii\ regions lie contain other \hii\
% regions whose \othreea\ oxygen abundances are consistent with lower
% branch values.
}
\label{fig_brightline}
\end{figure}

\begin{figure} % FIGURE 12
% \epsscale{0.5} % 0.8
% \plotone{fig12.ps}
\caption{
Direct (\othreea) oxygen abundances versus bright-line abundances 
for \hii\ regions in field and Virgo dIs.
Filled circles indicate \hii\ regions from field dIs.
VCC~0848-1 and VCC~1554-1 are indicated as crosses.
All symbols here represent \hii\ regions with \othreea\ abundances.
The solid line is the line of equality between direct and 
bright-line oxygen abundances.
The scatter of data points about the line of equality is consistent
with the $\approx$ 0.2~dex uncertainty (dotted lines) 
in the bright-line calibration.
}
\label{fig_oxydiff}
\end{figure}

\begin{figure} % FIGURE 13
% \epsscale{0.5} % 0.8
% \plotone{fig13.ps}
\caption{
Oxygen abundance versus absolute magnitude in $B$ for 
field and Virgo dIs.
Galaxy luminosity increases to the right.
The solid dots mark the field dIs.
The arrow indicates a lower limit to the oxygen abundance
for NGC~1560.
The solid line is a fit to the field dIs given by
Equation~(\ref{eqn_z_mb_fit}).
Crosses indicate Virgo dIs; gas-poor Virgo dIs are marked as crosses
enclosed by open circles (see \S~\ref{sec_z_mu}).
The error bars indicate typical uncertainties in the oxygen abundance
and absolute magnitude.
The uncertainty is at most 0.1~dex for oxygen abundances which
were determined directly from \othreea\ measurements.
A typical uncertainty of 0.2~mag in absolute magnitude accounts for
the various methods used to determine distances to field dIs 
and for possible uncertainties in the distances to Virgo dIs
arising from the ``depth'' of the Virgo Cluster.
}
\label{fig_z_mb}
\end{figure}

% \clearpage

\begin{figure} % FIGURE 14
% \epsscale{0.5} % 0.8
% \plotone{fig14.ps}
\caption{
$B-V$ color versus absolute magnitude in $B$ for 
field and Virgo dIs. 
Filled circles mark field dIs.
Gas-normal and gas-poor Virgo dIs (see \S~\ref{sec_z_mu}) are marked
by crosses and crosses enclosed by large open circles, respectively.
A fading vector is indicated by a solid arrow to show that
$B-V$ reddens by about $+$0.25~mag for each magnitude of fading
\citep{mb94}. 
}
\label{fig_bv_mb}
\end{figure}

% \clearpage

\begin{figure} % FIGURE 15
% \epsscale{0.5} % 0.8
% \plotone{fig15.ps}
\caption{
Oxygen abundance versus gas fraction for field and Virgo dIs.
The abundance increases upwards and gas fraction decreases
to the right.
The filled circles indicate dIs in the field.
The upward arrow indicates a lower limit to the oxygen abundance for 
NGC~1560.
The solid line is a fit to the field dIs given by
Equation~(\ref{eqn_z_mu_fit}).
Crosses mark Virgo dIs; gas-poor dIs are also marked with 
as crosses enclosed by open circles.
The error bar at the upper left indicates typical uncertainties of
0.1~dex in \othreea\ oxygen abundances 
and 0.1~dex in log log~(1/$\mu$). 
The latter uncertainty is derived from an estimated 0.05~mag
uncertainty in $B-V$, which affects the derivation of $M_{\ast}$,
and an estimated 20\% uncertainty in \hi\ gas mass 
(i.e., \citealp{hr86}, \citealp{hoffman87} for field and Virgo dIs,
respectively).
}
\label{fig_z_mu}
\end{figure}

\begin{figure} % FIGURE 16
% \epsscale{0.5} % 0.8
% \plotone{fig16.ps}
\caption{
Histograms of the gas-deficiency index (GDI) for field and
Virgo dIs.
Bins are 0.2~dex wide.
Gas deficiency increases to the right.
The histogram for the 22~dIs in the field sample is indicated
with a dashed line.
The histogram for the 12~dIs in the Virgo Cluster sample
is indicated with a solid line. 
Most of the field dIs are clustered around ${\rm GDI} \approx 0$;
seven Virgo dIs have similar GDIs to the field dwarfs, whereas
five have ${\rm GDI} \ga +0.8$.
}
\label{fig_gdi_hist}
\end{figure}

\begin{figure} % FIGURE 17
% \epsscale{0.5} % 0.8
% \plotone{fig17.ps}
\caption{
Nitrogen-to-oxygen ratio versus oxygen abundance for star-forming
dwarf galaxies.  
Filled circles indicate \hii\ regions in the sample of field dIs.
An upper limit to log(N/O) is shown for IC~1613 from \cite{talent80},
but see also \cite{lgh03} for an update.
For the Virgo sample, crosses indicate \hii\ regions from gas-normal
dIs, and crosses enclosed by open circles indicate \hii\ regions
from gas-poor dIs.
% Upper limits to log(N/O) are shown for \hii\ regions VCC~0512-3 
% and VCC~1249-1, because of low signal-to-noise for \ntwob.
Stars mark \hii\ regions in a variety of other dwarf galaxies,
including BCDs \citep{garnett90,ks96,vanzee97,it99}.
Two models for the production of nitrogen are shown \citep{vce93}:
primary plus secondary production of nitrogen is indicated
with a solid line, whereas a purely secondary origin for nitrogen is 
indicated with a dashed line. 
The solar value \citep{zsolar} is indicated with the ``${\odot}$''
symbol.
Typical errors in log(O/H) and log(N/O) are shown, assuming
\othreea\ temperatures are known.
}
\label{fig_no_oxy}
\end{figure}

% \clearpage

\begin{figure} % FIGURE 18
% \epsscale{1.}
% \plottwo{fig18a.ps}{fig18b.ps}
\caption{
Gas-normal and gas-poor dIs in the Virgo Cluster.
Left panel: 
Locations of dIs in the Virgo Cluster sample.
The positions of gas-normal (filled squares; ${\rm GDI} \la +0.8$) and
gas-poor dIs (open squares; ${\rm GDI} \ga +0.8$) from the present
sample are superposed on a greyscale X-ray map of the Virgo Cluster
from the ROSAT All-Sky-Survey.  
North is at the top and East is to the left.
The extended X-ray emission comes from the hot intracluster gas 
distributed throughout the cluster \citep{bohringer94}.
The large, dark spot is centred near the position of the
giant elliptical M~87.
A bright X-ray halo also surrounds M~86 to the west; halos of lesser
degree surround M~60 to the east and M~49 to the south.
Also indicated are foreground stars, Abell clusters (A1541, A1553) and
quasars (``QSO'') in the background, and Virgo Cluster member galaxies
with NGC and Messier catalog designations.
Right panel (PostScript file):
Gas-deficiency index (GDI) of dIs versus X-ray surface brightness
within the Virgo Cluster. 
Symbols are the same as those in the left panel.
The surface brightness for VCC~1585 may be contaminated by
the background cluster Abell~1560 at $z = 0.244$ \citep{bes94}.
Upper limits to the \hi\ mass for VCC~1200 and VCC~1448 correspond to
lower limits to their respective GDI values.
}
\label{fig_vdi_gasdef}
\end{figure}

% \clearpage

\begin{figure} % FIGURE 19
% \epsscale{0.55} % 0.8
% \plotone{fig19.ps}
\caption{
Fading diagrams for gas-poor Virgo dwarfs.
Open squares indicate gas-poor Virgo dIs with ${\rm GDI} \ga +0.8$.
Open circles and filled circles indicate non-nucleated dwarf
ellipticals (dEs) and nucleated dwarf ellipticals (dE,Ns), 
respectively.
``Transition'' dwarfs (dI/dE) and dIs with very low H~I content
\citep{hoffman87} are indicated as inverted-Ys.
With a fading model which assumes constant scale length,
fading vectors are shown as solid arrows.
(a) 
Exponential scale length, $r_0$, versus absolute magnitude in $B$.
The dotted lines bracket approximately the phase space occupied
by dIs and dEs/dSphs discussed by \citet[PT96]{pt96}.
Values of the exponential scale length are taken from
\cite{bc93}, except for VCC~1249 \citep{pt96}.
(b) 
Effective surface brightness, $\mu_{{\rm eff},B}$, versus absolute
magnitude in $B$ for gas-poor Virgo dwarfs. 
The symbols are the same as in the top panel.
Effective surface brightness values are obtained from \cite{bc93}.
}
\label{fig_fading}
\end{figure}

\begin{figure} % FIGURE 20
% \epsscale{0.5} % 0.8
% \plotone{fig20.ps}
\caption{
Stellar mass-to-light ratios in $B$ versus $B-V$.
Filled circles mark the sample of field dIs.
Virgo dIs are marked by crosses; gas-poor dIs are additionally shown
as crosses enclosed in open circles.
From the color-based models of \cite{belldejong01}, the upper and
lower solid curves represent ``0.7 times Salpeter'' and ``Salpeter''
models, respectively.
The remaining three models are Bruzual \& Charlot (BC) synthesis
models (at 40\% of solar metallicity) for a Salpeter IMF (short-dash
line), a scaled Salpeter IMF (long-dash line), and a modified Salpeter
IMF (dot-dash line); see \cite{belldejong01} for details.
}
\label{fig_mstarlb}
\end{figure}

\clearpage	% TABLES (use \tablenum{n} when all tables entered)

\begin{deluxetable}{ccccccccccc}
% \rotate
\tabletypesize{\tiny} % small 11pt; footnotesize 10pt; % scriptsize 8pt
\renewcommand{\arraystretch}{1.}
\tablecolumns{11}
\tablewidth{0pt}
\tablecaption{
Basic data for the sample of dwarf irregulars in the Virgo Cluster.
\label{table_vccdi}
}
\tablehead{
\colhead{Name} & \colhead{Alternate} & \colhead{Type} &
\colhead{$B_T$} & \colhead{$\mu_{0,B}$} & \colhead{$r_{0,B}$} & 
\colhead{$B\!-\!V$} & \colhead{\vsun} & \colhead{$F_{21}$} &
\colhead{$W_{50}$} & \colhead{$V_{\rm max}$} \\
\colhead{\#} & \colhead{Name} & & \colhead{(mag)} & 
\colhead{(mag $\Box^{-2}$)} & \colhead{(arcsec)} &
\colhead{(mag)} & \colhead{(km s$^{-1}$)} & \colhead{(Jy km s$^{-1}$)}
& \colhead{(km s$^{-1}$)} & \colhead{(km s$^{-1}$)} \\
\colhead{(1)} & \colhead{(2)} & \colhead{(3)} & \colhead{(4)} &
\colhead{(5)} & \colhead{(6)} & \colhead{(7)} & \colhead{(8)} &
\colhead{(9)} & \colhead{(10)} & \colhead{(11)}
}
\startdata
\objectname{VCC 0512} & UGC 7421$\,$\tablenotemark{a} & SBm & 
        15.30 & 23.67 & 16.22 & 0.49 & $+$153 & 3.694 & 108 & \nodata \\
\objectname{VCC 0848} & A1223+06 & dI/BCD & 
        14.73 & \nodata & \nodata & 0.42 & $+$1537 & 8.2 & 115 & 157 \\
\objectname{VCC 0888} & 8$^\circ$26 & dI & 
        15.00 & 23.15 & 10.47 & 0.61 & $+$1090 & 1.442 & 79 & \nodata \\
\objectname{VCC 1114} & UGC 7596$\,$\tablenotemark{b} & dI & 
        14.66 & 22.05 & 9.04 & 0.56 & $+$560 & 0.278 & 25 & \nodata \\
\objectname{VCC 1179} & IC 3412 & dI/BCD & 
        14.87 & 23.15 & 10.23 & 0.39 & $+$764 & 0.433 & 54 & \nodata \\
\objectname{VCC 1200} & IC 3416 & dI & 
        14.78 & 23.50 & 14.45 & 0.52 & $-$123 & $<$ 0.395 & 
	\nodata & \nodata \\
\objectname{VCC 1249} & UGC 7636 & dI/dSph & 
        14.26 & 24.07 & 31.8 & 0.50 & $+$276 & 0.167 & 70 & \nodata \\
\objectname{VCC 1448} & IC 3475 & dI/dSph & 
     13.93 & 23.04 & 26.91 & 0.76 & $+$2572 & $<$ 0.507 & \nodata & \nodata \\
\objectname{VCC 1554} & NGC 4532 & Sm & 
        12.15 & 19.77 & 19.39 & 0.39 & $+$2013 & 50.9 & 191 & 102 \\
\objectname{VCC 1585} & IC 3522$\,$\tablenotemark{c} & dI & 
        15.20 & 23.01 & 13.18 & 0.53 & $+$664 & 10.7 & 104 & 57 \\
\objectname{VCC 1789} & A1236+05 & dI & 
        15.07 & \nodata & \nodata & 0.72 & $+$1620 & 1.065 & 98 & \nodata \\
\objectname{VCC 2037} & 10$^\circ$71 & dI/BCD & 
        15.80 & 23.60 & 11.48 & 0.34 & $+$1142 & 0.358 & 36 & \nodata \\
\enddata
\tablenotetext{a}{Also known as 12\degr25.}
\tablenotetext{b}{Also known as 8\degr30.}
\tablenotetext{c}{Also known as DDO~136.}
\tablecomments{
Column (1): Galaxy number from Virgo Cluster Catalog \citep{vcc}.
Column (2): Alternate name to the galaxy.
Column (3): Morphological type.
Column (4): Total apparent $B$ magnitude.
Columns (5) and (6): Central surface brightness in $B$ and
exponential scale length in $B$ \citep{bc93,hoffman99,pt96}.
Column (7): \bv\ color \citep{bothun86,gh86}.
Column (8): Heliocentric velocity \citep{vcc,bpt93}.
Column (9): 21-cm \hi\ flux integral
\citep{hoffman87,hoffman96}. 
Column (10): \hi\ profile width at half-maximum
\citep{hoffman87}, except VCC~1249 \citep{pt92} and
VCC~1554 \citep{hoffman99}.
Column (11): \hi\ maximum rotational velocity:
VCC~0848 \citep{hoffman96}, VCC~1554 \citep{hoffman99},
VCC~1585 \citep{skillman87}.
}
\end{deluxetable}

% \clearpage 

\begin{deluxetable}{ccc}
\renewcommand{\arraystretch}{1.}
\tablecolumns{3}
\tabletypesize{\footnotesize} % small 11pt; footnotesize 10pt; scriptsize 8pt
\tablecaption{
Instrument configurations at KPNO and CFHT.
\label{table_obsprops}}
\tablewidth{0pt}
\tablehead{
\colhead{Property} & \colhead{KPNO 4-m} & \colhead{CFHT}
}
\startdata
Spectrograph & Ritchey-Chr\'etien (RCS) & Multi-Object (MOS) \\
CCD & Tektronix T2KB & STIS2 \\
Total area\tablenotemark{a} & 
	2048 pix $\times$ 2048 pix & 2048 pix $\times$ 2048 pix \\
Pixel size & 24 $\mu$m & 21 $\mu$m \\
Image scale & 0.69 arcsec pix$^{-1}$ & 0.44 arcsec pix$^{-1}$ \\
Gain & 3.1 $e^-$ ADU$^{-1}$ & 4.52 $e^-$ ADU$^{-1}$ \\
Read-noise & 4 $e^-$ (rms) & 9.3 $e^-$ (rms) \\
\tableline
Grating & KPC-10A & B600 \\
Groove density & 316 lines mm$^{-1}$ & 600 lines mm$^{-1}$ \\
Blaze $\lambda$ (1st order) & 4000 \AA & 5000 \AA \\
Dispersion & 2.75 \AA\ pixel$^{-1}$ & 2.24 \AA\ pixel$^{-1}$ \\
Spectral resolution\tablenotemark{b} & 6.9 \AA\ & 5.6 \AA \\
Effective $\lambda$ range & 
	3500--7600 \AA & 3600--7000 \AA\ \tablenotemark{c}\\
\tableline
Slit & Single long slit & Multi-slits \\
Length & 5.4 arcmin & 10 arcsec \\
Width & 2 arcsec & 2 arcsec \\
\tableline
Guide stars & \nodata & Hole diameters \\
Faint stars & \nodata & 3 arcsec (6.9 pix) \\
Bright stars & \nodata & 5 arcsec (11.5 pix) \\
\enddata
\tablenotetext{a}{
Wavelength coverage (1500 pix), spatial coverage (471 pix).
}
\tablenotetext{b}{
Based on 2.5 pixels FWHM corresponding to a slit width of
2 arcsec.
}
\tablenotetext{c}{
For slits placed at the centre of the field.}
% \tablecomments{}
\end{deluxetable}

% \clearpage

\begin{deluxetable}{ccccccccc}
\renewcommand{\arraystretch}{1.}
\tablecolumns{9}
\tabletypesize{\footnotesize} % small 11pt; footnotesize 10pt; scriptsize 8pt
\tablecaption{
Log of Observations.
\label{table_obslog}}
\tablewidth{0pt}
\tablehead{
\colhead{VCC \#} & \colhead{Obs.} & \colhead{Date} &
\colhead{$N_{\rm exp}$} & \colhead{$t_{\rm total}$} & 
\colhead{$\langle X \rangle$} & \colhead{$N_{\rm H II}$} &
\colhead{\othreea} & \colhead{RMS} \\
& & \colhead{(UT)} & & \colhead{(s)} & & & & \\
\colhead{(1)} & \colhead{(2)} & \colhead{(3)} & \colhead{(4)} &
\colhead{(5)} & \colhead{(6)} & \colhead{(7)} & \colhead{(8)} &
\colhead{(9)}
}
\startdata
0512 & CFHT & 1999 April 9 & 2$\times$1800 & 3600 & 1.27 &
        2 & upper limit & 5.0 \% \\
0848 & KPNO & 1997 March 3 & 4$\times$1800 & 7200 & 1.29 &
        1 & detected & 4.9 \% \\
0888 & KPNO & 1997 March 3 & 3$\times$1800 & 5400 & 1.10 &
        2 & upper limit & 4.9 \% \\
1114 & CFHT & 1999 April 9 & 2$\times$1800 & 3600 & 1.05 &
        2 & upper limit & 5.0 \% \\
1179 & KPNO & 1997 March 2 & 4$\times$1800 & 7200 & 1.10 &
        4 & upper limit & 5.5 \% \\
1200 & KPNO & 1997 March 3 & 3$\times$1800 & 5400 & 1.23 &
        1 & upper limit & 4.9 \% \\
1249 & CFHT & 1999 April 11 & 1$\times$1800 & 
	\phm{x}816\tablenotemark{a} & 1.03 &
        1 & \ldots\tablenotemark{b} & 4.3 \% \\
1448 & CFHT & 1999 April 11 & 3$\times$1800 & 5400 & 1.20 &
        0 & \ldots & 4.3 \% \\
1554 & CFHT & 1999 April 11 & 2$\times$1800 & 3600 & 1.05 &
        5 & detected & 4.3 \% \\
1585 & KPNO & 1997 March 2 & 2$\times$1800 & 3600 & 1.59 &
        1 & upper limit & 5.5 \% \\
1789 & KPNO & 1997 March 2 & 3$\times$1800 & 5400 & 1.30 &
        1 & upper limit & 5.5 \% \\
2037 & KPNO & 1997 March 2 & 3$\times$1800 & 5400 & 1.22 &
        2 & upper limit & 5.5 \% \\
\enddata
\tablenotetext{a}{1800 second exposure terminated due to increasing
cloud cover and rising humidity.}
\tablenotetext{b}{\hgamma\ was not detected in the spectrum.}
\tablecomments{
Observations obtained at KPNO and CFHT.
Column (1): Galaxy number from Virgo Cluster Catalog
\citep{vcc}.
Column (2): Observatory.
Column (3): Date of observation.
Column (4): Number of exposures obtained and the length
of each exposure.
Column (5): Total exposure time.
Column (6): Mean effective airmass.
Column (7): Number of \hii\ region spectra obtained.
Column (8): \othreea\ detection.
Column (9): Relative root-mean-square error in the
sensitivity function obtained from observations of standard stars
for the specified observing night.
}
\end{deluxetable}

% \clearpage

\begin{deluxetable}{ccccc}
\renewcommand{\arraystretch}{1.}
\tablecolumns{5}
\tabletypesize{\scriptsize} % small 11pt; footnotesize 10pt; scriptsize 8pt
\tablecaption{
Measured equivalent widths (EWs) at \hbeta\ for \hii\ regions in 
Virgo dwarfs.
\label{table_eqwidths}
}
\tablewidth{0pt}
\tablehead{
\colhead{VCC dI} & \colhead{EW of H$\beta$} & 
\colhead{EW of H$\beta$ emission in} & 
\colhead{EW of Total Absorption} & \colhead{Eff. EW of underlying} \\
\colhead{H~II Region} & \colhead{emission only,} & 
\colhead{simultaneous fit (\AA)} & \colhead{at H$\beta$ (\AA)} &
\colhead{absorption at H$\beta$,} \\
\colhead{} & \colhead{$W_{\rm e}$(\hbeta) (\AA)} & \colhead{} &
\colhead{} & \colhead{$W_{\rm abs}$(\hbeta) (\AA)} \\
\colhead{(1)} & \colhead{(2)} & \colhead{(3)} & \colhead{(4)} &
\colhead{(5)}
}
\startdata
0848-1 & $33.69 \pm 0.57$ & \nodata & \nodata & \nodata \\ 
0888-1 & $9.01 \pm 0.33$ & $10.40 \pm 0.40$ & 
         $4.99 \pm 0.94$ & $1.38 \pm 0.52$  \\ 
0888-2 & $6.59 \pm 0.31$ & $7.77 \pm 0.37$ &  
         $4.61 \pm 0.95$ & $1.19 \pm 0.49$ \\ 
1114-1 & $47.0 \pm 5.9$ & \nodata & \nodata & \nodata \\
1114-2 & $260 \pm 130$ & \nodata & \nodata & \nodata \\
1179-1 & $1.61 \pm 0.23$ & $3.73 \pm 0.21$ &  
        $5.44 \pm 0.42$ & $2.12 \pm 0.31$ \\ 
1179-2 & $6.09 \pm 0.28$ & $7.80 \pm 0.32$ & 
        $4.58 \pm 0.58$ & $1.72 \pm 0.43$ \\ 
1179-3 & $13.79 \pm 0.66$ & \nodata & \nodata & \nodata \\
1179-4 & $16.0 \pm 1.9$ & \nodata & \nodata & \nodata \\ 
1200-2 & $6.85 \pm 0.43$ & $9.52 \pm 0.51$ & 
        $7.3 \pm 1.0$ & $2.67 \pm 0.66$ \\ 
1249-1 & $26.8 \pm 5.9$ & \nodata & \nodata & \nodata \\ 
1554-1 & $112.3 \pm 6.8$ & \nodata & \nodata & \nodata \\
1554-2 & \nodata\tablenotemark{a} & \nodata & \nodata & \nodata \\
1554-3 & \nodata\tablenotemark{a} & \nodata & \nodata & \nodata \\
1554-4 & $4.33 \pm 0.33$ & \nodata & \nodata & \nodata \\
1554-5 & \nodata\tablenotemark{a} & \nodata & \nodata & \nodata \\
1585-1 & $66 \pm 13$ & \nodata & \nodata & \nodata \\
1789-1 & $6.18 \pm 0.18$ & $7.13 \pm 0.22$ & 
        $2.84 \pm 0.45$ & $0.94 \pm 0.29$ \\
2037-1 & $15.03 \pm 0.38$ & $16.27 \pm 0.49$ &  
        $4.29 \pm 0.93$ & $1.24 \pm 0.62$ \\
2037-2 & $3.85 \pm 0.47$ & $5.33 \pm 0.67$ &  
        $4.3 \pm 1.4$ & $1.48 \pm 0.82$ \\
\enddata
\tablenotetext{a}{Negative equivalent width, because the continuum is 
negative.}
\tablecomments{
Column~(1): \hii\ region.
Column~(2): Equivalent width of emission at \hbeta\ from a fit
of emission only (see Figure~\ref{fig_linefits}).
Column~(3): Equivalent width of the \hbeta\ emission obtained
from a simultaneous fit of absorption and emission 
(Figure~\ref{fig_linefits}).
Column~(4): Total equivalent width of the entire absorption
profile from the simultaneous fit.
Column~(5): Effective equivalent width of the underlying absorption
directly affecting \hbeta\ emission, which is the difference of
column~(3) and column~(2). 
}
\end{deluxetable}

% \clearpage

\begin{table}
\scriptsize	% small 11pt; footnotesize 10pt; scriptsize 8pt
\begin{center}
\renewcommand{\arraystretch}{1.}
\caption{
Observed and corrected line ratios for \hii\ regions in Virgo dwarfs.
\vspace*{3mm}
\label{table_allvirgo_data1}
}
\begin{tabular}{ccccccc}
\tableline \tableline
& \multicolumn{2}{c}{VCC 0512-3} & 
\multicolumn{2}{c}{VCC 0512-4} &
\multicolumn{2}{c}{VCC 0848-1} \\
\cline{2-7}
\multicolumn{1}{c}{Identification (\AA)} &
\multicolumn{1}{c}{$F$} & \multicolumn{1}{c}{$I$} &
\multicolumn{1}{c}{$F$} & \multicolumn{1}{c}{$I$} &
\multicolumn{1}{c}{$F$} & \multicolumn{1}{c}{$I$} \\ 
\tableline
$[\rm{O\;II}]\;3727$ &
        $263 \pm 76$ & $257 \pm 98$ &
        $458 \pm 131$ & $430 \pm 190$ &
        $252.1 \pm 1.2$ & $254 \pm 25$ 
\\
$[\rm{Ne\;III}]\;3869$ &
	\nodata & \nodata & \nodata & \nodata &
        $31.8 \pm 2.3$ & $31.7 \pm 4.4$
\\ 
${\rm H}\delta\;4101$ & 
	\nodata & \nodata & \nodata & \nodata &
        $15.8 \pm 1.3$ & $23.0 \pm 4.1$
\\ 
${\rm H}\gamma\;4340$ &
        $89 \pm 13$ & $86 \pm 22$ &
        $36 \pm 14$ & $43 \pm 30$ &
        $42.23 \pm 0.88$ & $47.2 \pm 5.1$
\\ 
$[{\rm O\;III}]\;4363$ &
        $<$ 25.2 (2$\sigma$) & $<$ 24.7 (2$\sigma$) & 
	\nodata & \nodata &
        $4.71 \pm 0.70$ & $4.55 \pm 0.95$
\\
He I 4472 & 
	\nodata & \nodata & \nodata & \nodata &
        $2.77 \pm 0.45$ & $2.66 \pm 0.59$
\\
${\rm H}\beta\;4861$ &
        $100.0 \pm 6.5$ & $100.0 \pm 8.6$ &
        $100 \pm 13$ & $100 \pm 15$ &
        $100 \pm 1.2$ & $100.0 \pm 5.1$
\\ 
$[{\rm O\;III}]\;4959$ &
        $179.3 \pm 7.4$ & $175 \pm 26$ &
        $93 \pm 13$ & $87 \pm 26$ &
        $124.9 \pm 1.2$ & $117 \pm 12$
\\
$[{\rm O\;III}]\;5007$ &
        $528.4 \pm 9.2$ & $517 \pm 72$ &
        $208 \pm 15$ & $193 \pm 48$ &
        $374.5 \pm 1.5$ & $352 \pm 35$
\\
$[{\rm N\;I}]\;5198$ & 
	\nodata & \nodata & \nodata & \nodata &
        $1.73 \pm 0.42$ & $1.61 \pm 0.48$
\\
He I 5876 & 
	\nodata & \nodata & \nodata & \nodata &
        $11.81 \pm 0.98$ & $10.8 \pm 1.6$
\\ 
$[{\rm O\;I}]\;6300$ & 
	\nodata & \nodata & \nodata & \nodata &
        $7.26 \pm 0.93$ & $6.6 \pm 1.2$
\\
$[{\rm S\;III}]\;6312$ & 
	\nodata & \nodata & \nodata & \nodata &
        $3.29 \pm 0.75$ & $2.98 \pm 0.84$
\\
$[{\rm O\;I}]\;6364$ & 
	\nodata & \nodata & \nodata & \nodata &
        $2.29 \pm 0.73$ & $2.07 \pm 0.78$
\\
$[{\rm N\;II}]\;6548$ &
        $2.3 \pm 4.1$ & $2.3 \pm 4.3$ & \nodata & \nodata &
        $7.5 \pm 2.0$ & $6.7 \pm 2.2$ 
\\
${\rm H}\alpha\;6563$ &
        $265.1 \pm 5.2$ & $260 \pm 37$ &
        $270 \pm 12$ & $255 \pm 58$ &
        $314.3 \pm 2.5$ & $286 \pm 29$
\\ 
$[\rm{N\;II}]\;6583$ &
        $6.2 \pm 4.2$ & $6.1 \pm 4.6$ &
        $24.1 \pm 9.1$ & $22 \pm 12$ &
        $20.8 \pm 2.0$ & $18.7 \pm 3.0$
\\
He I 6678 & 
	\nodata & \nodata & \nodata & \nodata &
        $2.95 \pm 0.39$ & $2.65 \pm 0.51$
\\ 
$[{\rm S\;II}]\;6716$ &
        $33.1 \pm 3.8$ & $32.4 \pm 6.8$ & \nodata & \nodata &
        $34.13 \pm 0.46$ & $30.7 \pm 3.1$
\\ 
$[{\rm S\;II}]\;6731$ &
        $24.5 \pm 3.7$ & $23.9 \pm 5.9$ & \nodata & \nodata &
        $24.11 \pm 0.44$ & $21.6 \pm 2.2$
\\ 
He I 7065 &
	\nodata & \nodata & \nodata & \nodata &
        $2.46 \pm 0.43$ & $2.20 \pm 0.51$
\\ 
$[{\rm Ar\;III}]\;7136$ &
	\nodata & \nodata & \nodata & \nodata &
        $7.65 \pm 0.52$ & $6.83 \pm 0.92$
\\ 
\tableline
& \multicolumn{2}{c}{VCC 0888-1} & 
\multicolumn{2}{c}{VCC 0888-2} &
\multicolumn{2}{c}{VCC 1114-1} \\
\cline{2-7}
\multicolumn{1}{c}{Identification (\AA)} &
\multicolumn{1}{c}{$F$} & \multicolumn{1}{c}{$I$} &
\multicolumn{1}{c}{$F$} & \multicolumn{1}{c}{$I$} &
\multicolumn{1}{c}{$F$} & \multicolumn{1}{c}{$I$} \\ 
\tableline
$[\rm{O\;II}]\;3727$ &
        $382 \pm 21$ & $394 \pm 51$ &
        $427 \pm 22$ & $475 \pm 71$ &
        $419 \pm 98$ & $400 \pm 130$ 
\\
${\rm H}\gamma\;4340$ &
        $30.1 \pm 4.4$ & $46 \pm 13$ &
        $27.0 \pm 3.3$ & $50 \pm 14$ &
        $45 \pm 10$ & $49 \pm 18$
\\
$[{\rm O\;III}]\;4363$ &
        $<$ 5.36 (2$\sigma$) & $<$ 4.76 (2$\sigma$) &
        $<$ 4.70 (2$\sigma$) & $<$ 4.12 (2$\sigma$) & 
        \nodata & \nodata
\\
${\rm H}\beta\;4861$ &
        $100.0 \pm 6.5$ & $100.0 \pm 7.4$ &
        $100.0 \pm 8.1$ & $100.0 \pm 9.7$ &
        $100.0 \pm 7.4$ & $100.0 \pm 9.5$ 
\\
$[{\rm O\;III}]\;4959$ &
        $46.5 \pm 4.2$ & $37.5 \pm 6.2$ &
        $89.3 \pm 6.5$ & $67 \pm 11$ &
        $20.6 \pm 5.9$ & $19.8 \pm 7.7$
\\
$[{\rm O\;III}]\;5007$ &
       $138.4 \pm 8.0$ & $111 \pm 15$ &
        $287 \pm 15$ & $213 \pm 32$ &
        $91.2 \pm 8.0$ & $87 \pm 17$
\\
$[{\rm N\;II}]\;6548$ &
        $12.9 \pm 4.0$ & $9.0 \pm 3.4$ & 
        $11.0 \pm 5.7$ & $6.5 \pm 4.0$ &
        $15.4 \pm 4.8$ & $14.8 \pm 6.1$
\\
${\rm H}\alpha\;6563$ &
        $396 \pm 20$ & $286 \pm 36$ & 
        $467 \pm 24$ & $286 \pm 43$ &
        $291.4 \pm 6.0$ & $278 \pm 42$ 
\\
$[\rm{N\;II}]\;6583$ &
        $52.9 \pm 4.7$ & $36.6 \pm 6.0$ &
        $56.7 \pm 6.3$ & $33.3 \pm 6.9$ &
        $41.3 \pm 4.9$ & $39.6 \pm 8.9$
\\
$[{\rm S\;II}]\;6716$ &
        $78.3 \pm 5.0$ & $53.7 \pm 7.4$ &
        $90.3 \pm 5.2$ & $52.2 \pm 8.1$ &
        $38.2 \pm 5.4$ & $36.7 \pm 9.0$ 
\\
$[{\rm S\;II}]\;6731$ &
        $59.5 \pm 4.2$ & $40.8 \pm 6.0$ &
        $64.2 \pm 4.0$ & $37.1 \pm 5.9$ &
        $27.8 \pm 5.1$ & $26.7 \pm 7.7$ 
\\
$[{\rm Ar\;III}]\;7136$ &
        \nodata & \nodata &
        $15.3 \pm 3.7$ & $8.5 \pm 2.9$ &
        \nodata & \nodata
\\ 
\tableline
\end{tabular}
\tablecomments{
Emission lines are listed in \AA.
$F$ is the observed flux ratio with respect to \hbeta.
$I$ is the corrected intensity ratio, corrected for underlying
Balmer absorption and the adopted reddening listed in
Table~\ref{table_vccspecprops}.
The uncertainties in the observed line ratios account for the
uncertainties in the fits to the line profiles, the surrounding
continua, and the relative uncertainty in the sensitivity function
listed in Table~\ref{table_obslog}. 
Flux uncertainties in the \hbeta\ reference line are not included.
Uncertainties in the corrected line ratios account for uncertainties
in the specified line and in the \hbeta\ reference line.
}
\end{center}
\end{table}

% \clearpage

\begin{table}
\scriptsize	% small 11pt; footnotesize 10pt; scriptsize 8pt
\begin{center}
\renewcommand{\arraystretch}{1.}
\caption{
Observed and corrected line ratios for \hii\ regions in Virgo
dwarfs (continued). 
\vspace*{3mm}
\label{table_allvirgo_data2}
}
\begin{tabular}{ccccccc}
\tableline \tableline
& \multicolumn{2}{c}{VCC 1114-2} & 
\multicolumn{2}{c}{VCC 1179-1} &
\multicolumn{2}{c}{VCC 1179-2} \\
\cline{2-7}
\multicolumn{1}{c}{Identification (\AA)} &
\multicolumn{1}{c}{$F$} & \multicolumn{1}{c}{$I$} &
\multicolumn{1}{c}{$F$} & \multicolumn{1}{c}{$I$} &
\multicolumn{1}{c}{$F$} & \multicolumn{1}{c}{$I$} \\ 
\tableline
$[\rm{O\;II}]\;3727$ &
        $304 \pm 63$ & $302 \pm 92$ &
        $690 \pm 65$ & $528 \pm 168$ &
        $506 \pm 32$ & $496 \pm 71$ 
\\
$[\rm{Ne\;III}]\;3869$ &
        \nodata & \nodata &
        $110 \pm 43$ & $77 \pm 48$&
        $36 \pm 12$ & $34 \pm 14$
\\
${\rm H}\gamma\;4340$ &
        $33.7 \pm 8.1$ & $36 \pm 13$ &
        \nodata & \nodata &
        \nodata & \nodata
\\
$[{\rm O\;III}]\;4363$ &
        \nodata & \nodata &
        \nodata & \nodata &
        $<$ 4.2 (2$\sigma$) & $<$ 3.5 (2$\sigma$) 
\\
${\rm H}\beta\;4861$ &
        $100.0 \pm 7.3$ & $100.0 \pm 9.3$ &
        $100 \pm 14$ & $100 \pm 23$ &
        $100.0 \pm 6.9$  & $100.0 \pm 8.0$ 
\\
$[{\rm O\;III}]\;4959$ &
        \nodata & \nodata &
        $153 \pm 16$ & $69 \pm 23$ &
        $70.6 \pm 5.6$ & $52.2 \pm 8.3$ 
\\
$[{\rm O\;III}]\;5007$ &
        $29.0 \pm 6.3$ & $28.8 \pm 9.1$ &
        $468 \pm 30$ & $207 \pm 59$ &
        $207 \pm 12$ & $152 \pm 21$
\\
He I 5876 &
        \nodata & \nodata &
        $57.2 \pm 8.8$ & $20.6 \pm 7.8$ &
        $18.5 \pm 3.9$ & $12.2 \pm 3.6$ 
\\
$[{\rm N\;II}]\;6548$ &
        $9.5 \pm 5.1$ & $9.4 \pm 6.0$ &
        $35 \pm 10$ & $11.1 \pm 5.9$ &
        $17.3 \pm 2.9$ & $10.8 \pm 2.7$ 
\\
${\rm H}\alpha\;6563$ &
        $248.2 \pm 6.3$ & $247 \pm 37$ &
        $804 \pm 46$ & $286 \pm 81$ &
        $438 \pm 25$ & $286 \pm 39$
\\
$[\rm{N\;II}]\;6583$ &
        $31.4 \pm 5.1$ & $31.2 \pm 8.2$ &
        $105 \pm 12$ & $34 \pm 11$ &
        $67.0 \pm 4.6$ & $41.6 \pm 6.2$ 
\\
He I 6678 &
        \nodata & \nodata &
        \nodata & \nodata &
        $5.5 \pm 2.1$ & $3.4 \pm 1.6$
\\
$[{\rm S\;II}]\;6716$ &
        $47.2 \pm 4.5$ & $46.9 \pm 9.4$ &
        $136.0 \pm 9.4$ & $43 \pm 12$ &
        $98.5 \pm 6.0$ & $60.6 \pm 8.5$ 
\\
$[{\rm S\;II}]\;6731$ &
        $31.8 \pm 4.2$ & $31.5 \pm 7.4$ &
        $105.8 \pm 8.0$ & $33.2 \pm 9.9$ &
        $66.4 \pm 4.3$ & $40.8 \pm 5.9$
\\
$[{\rm Ar\;III}]\;7136$ &
        \nodata & \nodata &
        $38 \pm 10$ & $11.4 \pm 5.7$ &
        $12.7 \pm 3.5$ & $7.6 \pm 2.7$
\\ 
\tableline
& \multicolumn{2}{c}{VCC 1179-3} & 
\multicolumn{2}{c}{VCC 1179-4} &
\multicolumn{2}{c}{VCC 1200-1\tablenotemark{a}} \\
\cline{2-7}
\multicolumn{1}{c}{Identification (\AA)} &
\multicolumn{1}{c}{$F$} & \multicolumn{1}{c}{$I$} &
\multicolumn{1}{c}{$F$} & \multicolumn{1}{c}{$I$} &
\multicolumn{1}{c}{$F$} & \multicolumn{1}{c}{$I$} \\ 
\tableline
$[\rm{O\;II}]\;3727$ &
        $254 \pm 18$ & $222 \pm 33$ &
        $334 \pm 25$ & $329 \pm 69$ &
        $114.1 \pm 8.0$ & \nodata
\\
$[\rm{Ne\;III}]\;3869$ &
        $38.7 \pm 9.4$ & $34 \pm 11$ &
        $13 \pm 13$ & $13 \pm 15$ &
        \nodata & \nodata
\\ 
${\rm H}\gamma\;4340$ &
        $47.1 \pm 5.7$ & $53 \pm 12$ &
        $30.9 \pm 7.7$ & $38 \pm 18$ &
        \nodata & \nodata
\\
$[{\rm O\;III}]\;4363$ &
        $<$ 7.7 (2$\sigma$) & $<$ 6.7 (2$\sigma$) &
        $<$ 14 (2$\sigma$) & $<$ 13 (2$\sigma$) &
        \nodata & \nodata
\\
${\rm H}\beta\;4861$ &
        $100.0 \pm 7.1$ & $100.0 \pm 7.7$ &
        $100 \pm 12$ & $100 \pm 13$ &
        \nodata & \nodata
\\ 
$[{\rm O\;III}]\;4959$ &
        $116.3 \pm 8.1$ & $102 \pm 15$ &
        $70.7 \pm 9.0$ & $62 \pm 16$ &
        \nodata & \nodata
\\
$[{\rm O\;III}]\;5007$ &
        $335 \pm 19$ & $293 \pm 40$ &
        $245 \pm 17$ & $216 \pm 44$ &
        $15.5 \pm 2.8$ & \nodata
\\
$[{\rm N\;II}]\;6548$ &
        $9.2 \pm 5.3$ & $8.1 \pm 5.3$ &
        $9.7 \pm 9.2$ & $8.0 \pm 8.8$ &
        \nodata & \nodata
\\ 
${\rm H}\alpha\;6563$ &
        $299 \pm 18$ & $269 \pm 37$ &
        $337 \pm 22$ & $286 \pm 58$ &
        $100 \pm 4$ & \nodata
\\
$[\rm{N\;II}]\;6583$ &
        $15.4 \pm 5.4$ & $13.4 \pm 5.8$ &
        $37.3 \pm 9.5$ & $31 \pm 12$ &
        $8.2 \pm 3.2$ & \nodata
\\
$[{\rm S\;II}]\;6716$ &
        $28.6 \pm 4.6$ & $25.0 \pm 5.9$ &
        $47.4 \pm 7.8$ & $39 \pm 12$ &
        $32.6 \pm 2.9$ & \nodata
\\ 
$[{\rm S\;II}]\;6731$ &
        $14.1 \pm 3.6$ & $12.3 \pm 4.1$ &
        $28.7 \pm 6.3$ & $23.6 \pm 8.4$ &
        $26.4 \pm 2.8$ & \nodata
\\ 
$[{\rm Ar\;III}]\;7136$ &
        $12.7 \pm 4.8$ & $11.1 \pm 5.1$ &
        $35.0 \pm 8.7$ & $29 \pm 11$ &
        \nodata & \nodata
\\
\tableline
\end{tabular}
\tablenotetext{a}{
\hbeta\ and \othreeb\ lines were not detected.
Observed flux ratios are not corrected and are given with respect to
\halpha, where $F$(\halpha) = 
$(1.16 \pm 0.05) \times 10^{-15}$~ergs~s$^{-1}$~cm$^{-2}$.
}
\tablecomments{
See Table~\ref{table_allvirgo_data1}.
}
\end{center}
\end{table}

% \clearpage

\begin{table}
\scriptsize	% small 11pt; footnotesize 10pt; scriptsize 8pt
\begin{center}
\renewcommand{\arraystretch}{1.}
\caption{
Observed and corrected line ratios for \hii\ regions in Virgo
dwarfs (continued). 
\vspace*{3mm}
\label{table_allvirgo_data3}
}
\begin{tabular}{ccccccc}
\tableline \tableline
& \multicolumn{2}{c}{VCC 1200-2} & 
\multicolumn{2}{c}{VCC 1249-1\tablenotemark{a}} &
\multicolumn{2}{c}{VCC 1554-1} \\
\cline{2-7}
\multicolumn{1}{c}{Identification (\AA)} &
\multicolumn{1}{c}{$F$} & \multicolumn{1}{c}{$I$} &
\multicolumn{1}{c}{$F$} & \multicolumn{1}{c}{$I$} &
\multicolumn{1}{c}{$F$} & \multicolumn{1}{c}{$I$} \\ 
\tableline
$[\rm{O\;II}]\;3727$ &
        $298 \pm 16$ & $246 \pm 33$ & 
        $693 \pm 166$ & $645 \pm 197$ &
        $474.0 \pm 6.8$ & $446 \pm 44$ 
\\
$[\rm{Ne\;III}]\;3869$ &
        \nodata & \nodata &
        \nodata & \nodata &
        $19.0 \pm 1.3$ & $18.7 \pm 2.4$ 
\\
He II 4026 &
        \nodata & \nodata &
        \nodata & \nodata &
        $2.05 \pm 0.68$ & $2.01 \pm 0.77$ 
\\
$[\rm{S\;II}]\;4069$ &
        \nodata & \nodata &
        \nodata & \nodata &
        $2.26 \pm 0.58$ & $2.22 \pm 0.69$ 
\\
${\rm H}\delta\;4101$ &
        \nodata & \nodata &
        \nodata & \nodata &
        $22.82 \pm 0.74$ & $25.3 \pm 2.7$ 
\\
${\rm H}\gamma\;4340$ &
        \nodata & \nodata &
        \nodata & \nodata &
        $47.8 \pm 1.1$ & $49.7 \pm 4.9$ 
\\
$[{\rm O\;III}]\;4363$ &
        \nodata & \nodata &
        \nodata & \nodata &
        $1.54 \pm 0.86$ & $1.51 \pm 0.92$ 
\\
He I 4472 &
        \nodata & \nodata &
        \nodata & \nodata &
        $2.94 \pm 0.48$ & $2.89 \pm 0.63$ 
\\
${\rm H}\beta\;4861$ &
        $100.0 \pm 7.1$ & $100.0 \pm 8.1$ &
        $100 \pm 17$ & $100 \pm 19$ &
        $100.0 \pm 2.0$ & $100.0 \pm 4.9$ 
\\
He I 4922 &
        \nodata & \nodata &
        \nodata & \nodata &
        $0.83 \pm 0.25$ & $0.81 \pm 0.29$ 
\\
$[{\rm O\;III}]\;4959$ &
        $38.0 \pm 3.6$ & $31.4 \pm 5.5$ &
        $84 \pm 17$ & $79 \pm 22$ &
        $78.4 \pm 3.2$ & $77.0 \pm 8.3$ 
\\ 
$[{\rm O\;III}]\;5007$ &
        $111.6 \pm 6.5$ & $92 \pm 13$ &
        $130 \pm 22$ & $121 \pm 31$ &
        $230.8 \pm 4.0$ & $227 \pm 22$ 
\\
$[{\rm N\;I}]\;5198$ & 
        \nodata & \nodata &
        \nodata & \nodata &
        $1.57 \pm 0.24$ & $1.55 \pm 0.32$ 
\\
$[{\rm N\;II}]\;5755$ &
        \nodata & \nodata &
        \nodata & \nodata &
        $0.52 \pm 0.17$ & $0.51 \pm 0.19$ 
\\
He I 5876 &
        $13.8 \pm 3.9$ & $11.4 \pm 4.2$ &
        \nodata & \nodata &
        $12.42 \pm 0.30$ & $12.2 \pm 1.2$ 
\\
$[{\rm O\;I}]\;6300$ & 
        \nodata & \nodata &
        \nodata & \nodata &
        $4.44 \pm 0.28$ & $4.36 \pm 0.55$ 
\\ 
$[{\rm S\;III}]\;6312$ & 
        \nodata & \nodata &
        \nodata & \nodata &
        $1.12 \pm 0.22$ & $1.10 \pm 0.28$ 
\\
$[{\rm O\;I}]\;6364$ & 
        \nodata & \nodata &
        \nodata & \nodata &
        $1.78 \pm 0.22$ & $1.75 \pm 0.32$ 
\\
$[{\rm N\;II}]\;6548$ &
        $5.4 \pm 3.4$ & $4.4 \pm 3.1$ &
        \nodata & \nodata &
        $6.4 \pm 2.6$ & $6.2 \pm 2.8$ 
\\ 
${\rm H}\alpha\;6563$ &
        $287 \pm 15$ & $247 \pm 33$ &
        $289 \pm 16$ & $270 \pm 54$ &
        $279.9 \pm 3.2$ & $276 \pm 26$ 
\\
$[\rm{N\;II}]\;6583$ &
        $22.2 \pm 3.5$ & $18.3 \pm 4.4$ &
        $13 \pm 13$ & $12 \pm 12$ &
        $21.1 \pm 2.6$ & $20.7 \pm 3.7$ 
\\ 
He I 6678 &
        \nodata & \nodata &
        \nodata & \nodata &
        $2.96 \pm 0.39$ & $2.91 \pm 0.54$ 
\\
$[{\rm S\;II}]\;6716$ &
        $57.0 \pm 4.3$ & $47.1 \pm 7.4$ &
        \nodata & \nodata &
        $25.64 \pm 0.42$ & $25.2 \pm 2.4$ 
\\
$[{\rm S\;II}]\;6731$ &
        $39.5 \pm 3.7$ & $32.6 \pm 5.7$ &
        \nodata & \nodata &
        $17.61 \pm 0.41$ & $17.3 \pm 1.7$ 
\\
He I 7065 &
        \nodata & \nodata &
        \nodata & \nodata &
        $1.69 \pm 0.25$ & $1.66 \pm 0.34$ 
\\
$[{\rm Ar\;III}]\;7136$ &
        \nodata & \nodata &
        \nodata & \nodata &
        $4.67 \pm 0.32$ & $4.59 \pm 0.60$ 
\\
\tableline
& \multicolumn{2}{c}{VCC 1554-2} & 
\multicolumn{2}{c}{VCC 1554-3} &
\multicolumn{2}{c}{VCC 1554-4} \\
\cline{2-7}
\multicolumn{1}{c}{Identification (\AA)} &
\multicolumn{1}{c}{$F$} & \multicolumn{1}{c}{$I$} &
\multicolumn{1}{c}{$F$} & \multicolumn{1}{c}{$I$} &
\multicolumn{1}{c}{$F$} & \multicolumn{1}{c}{$I$} \\ 
\tableline
$[\rm{O\;II}]\;3727$ &
        $330 \pm 18$ & $400 \pm 51$ &
        $283 \pm 24$ & $295 \pm 47$ &
        $595 \pm 39$ & $610 \pm 120$ 
\\
$[\rm{Ne\;III}]\;3869$ &
        \nodata & \nodata &
        $38.3 \pm 9.6$ & $40 \pm 13$ &
        \nodata & \nodata 
\\
${\rm H}\delta\;4101$ & 
        $46.2 \pm 6.2$ & $43 \pm 10$ &
        $39.9 \pm 5.6$ & $36.6 \pm 8.6$ &
        \nodata & \nodata 
\\
${\rm H}\gamma\;4340$ &
        $57.2 \pm 5.1$ & $54 \pm 10$ &
        $62.2 \pm 5.3$ & $60 \pm 10$ &
        $21.1 \pm 8.1$ & $73 \pm 68$ 
\\
$[{\rm O\;III}]\;4363$ &
        $<$ 9.8 (2$\sigma$) & $<$ 11 (2$\sigma$) &
        $<$ 13  (2$\sigma$) & $<$ 13 (2$\sigma$) &
        $<$ 11 (2$\sigma$) & $<$ 8.5 (2$\sigma$) 
\\
${\rm H}\beta\;4861$ &
        $100.0 \pm 3.2$ & $100.0 \pm 5.7$ &
        $100.0 \pm 4.5$ & $100.0 \pm 6.5$ &
        $100.0 \pm 7.6$ & $100 \pm 11$ 
\\
$[{\rm O\;III}]\;4959$ &
        $57.8 \pm 4.1$ & $61.9 \pm 8.6$ &
        $77.2 \pm 3.4$ & $81 \pm 10$ &
        $145.5 \pm 8.5$ & $97 \pm 18$ 
\\
$[{\rm O\;III}]\;5007$ &
        $178.7 \pm 5.0$ & $191 \pm 21$ &
        $241.3 \pm 4.3$ & $252 \pm 28$ &
        $399 \pm 10$ & $264 \pm 43$ 
\\ 
He I 5876 &
        $10.7 \pm 1.7$ & $10.9 \pm 2.4$ &
        $12.1 \pm 1.9$ & $12.6 \pm 2.9$ &
        $24.4 \pm 3.4$ & $13.7 \pm 3.5$ 
\\
$[{\rm N\;II}]\;6548$ &
        $10.2 \pm 3.3$ & $10.2 \pm 3.9$ &
        $4.8 \pm 2.8$ & $5.0 \pm 3.3$ &
        $22.2 \pm 8.7$ & $11.4 \pm 5.8$ 
\\
${\rm H}\alpha\;6563$ &
        $293.5 \pm 4.2$ & $286 \pm 29$ &
        $262.8 \pm 3.5$ & $270 \pm 30$ &
        $531 \pm 11$ & $286 \pm 46$ 
\\
$[\rm{N\;II}]\;6583$ &
        $29.6 \pm 3.4$ & $29.3 \pm 5.2$ &
        $12.0 \pm 2.8$ & $12.5 \pm 3.8$ &
        $55.0 \pm 8.8$ & $28.2 \pm 7.8$ 
\\
He I 6678 &
        $5.4 \pm 1.3$ & $5.4 \pm 1.6$ &
        \nodata & \nodata &
        \nodata & \nodata 
\\ 
$[{\rm S\;II}]\;6716$ &
        $21.4 \pm 1.5$ & $21.1 \pm 2.9$ &
        $45.8 \pm 7.9$ & $48 \pm 12$ &
        $92.5 \pm 4.3$ & $46.7 \pm 8.2$ 
\\
$[{\rm S\;II}]\;6731$ &
        $18.1 \pm 1.4$ & $17.9 \pm 2.6$ &
        $5.8 \pm 8.0$ & $5.8 \pm 8.8$ &
        $67.8 \pm 4.0$ & $34.2 \pm 6.3$ 
\\
\tableline
\end{tabular}
\tablenotetext{a}{
The \hii\ region, VCC~1249-1, is also labelled LR1; 
$F$ and $I$ values were previously reported in \cite{lrm00}.
}
\tablecomments{
See Table~\ref{table_allvirgo_data1}.
}
\end{center}
\end{table}

% \clearpage

\begin{table}
\scriptsize	% small 11pt; footnotesize 10pt; scriptsize 8pt
\begin{center}
\renewcommand{\arraystretch}{1.}
\caption{
Observed and corrected line ratios for \hii\ regions in Virgo
dwarfs (continued). 
\vspace*{3mm}
\label{table_allvirgo_data4}
}
\begin{tabular}{ccccccc}
\tableline \tableline
& \multicolumn{2}{c}{VCC 1554-5} & 
\multicolumn{1}{c}{VCC 1554-6\tablenotemark{a}} &
\multicolumn{1}{c}{VCC 1554-7\tablenotemark{a}} &
\multicolumn{2}{c}{VCC 1585-1} \\
\cline{2-7}
\multicolumn{1}{c}{Identification (\AA)} &
\multicolumn{1}{c}{$F$} & \multicolumn{1}{c}{$I$} &
\multicolumn{1}{c}{$F$} & \multicolumn{1}{c}{$F$ / $I$} &
\multicolumn{1}{c}{$F$} & \multicolumn{1}{c}{$I$} \\ 
\tableline
$[\rm{O\;II}]\;3727$ &
        $455 \pm 44$ & $496 \pm 99$ &
        $481 \pm 29$ & $258 \pm 40$ / $183 \pm 80$ &
        $338 \pm 28$ & $375 \pm 83$ 
\\
${\rm H}\gamma\;4340$ &
        $49.9 \pm 6.5$ & $46 \pm 12$ &
        \nodata & \nodata &
        $53.3 \pm 7.8$ & $59 \pm 17$ 
\\
$[{\rm O\;III}]\;4363$ &
        $<$ 16 (2$\sigma$) & $<$ 16 (2$\sigma$) &
        \nodata & \nodata &
        $<$ 16 (2$\sigma$) & $<$ 17 (2$\sigma$) 
\\
${\rm H}\beta\;4861$ &
        $100.0 \pm 7.3$ & $100.0 \pm 9.3$ &
        $100 \pm 12$ & $100 \pm 18$ / $100 \pm 28$ & 
        $100.0 \pm 9.6$ & $100 \pm 12$ 
\\
$[{\rm O\;III}]\;4959$ &
        $42.6 \pm 4.6$ & $46.5 \pm 9.8$ &
        $23 \pm 13$ & \nodata &
        $50.6 \pm 8.4$ & $49 \pm 14$ 
\\ 
$[{\rm O\;III}]\;5007$ &
        $150.2 \pm 5.9$ & $164 \pm 25$ &
        $126 \pm 17$ & \nodata &
        $120.8 \pm 9.9$ & $116 \pm 25$ 
\\
$[{\rm N\;II}]\;6548$ &
        $7.6 \pm 3.9$ & $8.3 \pm 5.1$ &
        \nodata & $85 \pm 18$ / $46 \pm 23$ &
        $24 \pm 13$ & $22 \pm 14$ 
\\
${\rm H}\alpha\;6563$ &
        $240.2 \pm 4.9$ & $255 \pm 36$ &
        \nodata & $451 \pm 22$ / $286 \pm 99$ & 
        $324 \pm 16$ & $286 \pm 55$ 
\\
$[\rm{N\;II}]\;6583$ &
        $17.8 \pm 3.9$ & $19.4 \pm 6.2$ &
        \nodata & $235 \pm 20$ / $127 \pm 47$ &
        $34 \pm 13$ & $30 \pm 15$ 
\\
$[{\rm S\;II}]\;6716$ &
        $41.4 \pm 3.1$ & $45.1 \pm 8.1$ &
        \nodata & \nodata &
        $43.4 \pm 8.4$ & $38 \pm 12$ 
\\ 
$[{\rm S\;II}]\;6731$ &
        $26.5 \pm 2.8$ & $28.9 \pm 6.0$ &
        \nodata & \nodata &
        $57.3 \pm 8.8$ & $50 \pm 14$ 
\\
\tableline
& \multicolumn{2}{c}{VCC 1789-1} & 
\multicolumn{2}{c}{VCC 2037-1} &
\multicolumn{2}{c}{VCC 2037-2} \\
\cline{2-7}
\multicolumn{1}{c}{Identification (\AA)} &
\multicolumn{1}{c}{$F$} & \multicolumn{1}{c}{$I$} &
\multicolumn{1}{c}{$F$} & \multicolumn{1}{c}{$I$} &
\multicolumn{1}{c}{$F$} & \multicolumn{1}{c}{$I$} \\ 
\tableline
$[\rm{O\;II}]\;3727$ &
        $404 \pm 23$ & $482 \pm 61$ &
        $403 \pm 23$ & $433 \pm 54$ &
        $492 \pm 41$ & $450 \pm 130$ 
\\
$[\rm{Ne\;III}]\;3869$ &
        \nodata & \nodata &
        $31.4 \pm 5.2$ & $32.6 \pm 7.6$ &
        \nodata & \nodata 
\\
${\rm H}\gamma\;4340$ &
        $30.5 \pm 4.7$ & $43 \pm 12$ &
        $37.1 \pm 3.6$ & $49.9 \pm 9.5$ &
        \nodata & \nodata 
\\
$[{\rm O\;III}]\;4363$ &
        $<$ 9.3 (2$\sigma$) & $<$ 9.0 (2$\sigma$) &
        $<$ 5.1 (2$\sigma$) & $<$ 4.8 (2$\sigma$) &
        \nodata & \nodata 
\\
${\rm H}\beta\;4861$ &
        $100.0 \pm 6.4$ & $100.0 \pm 6.9$ &
        $100.0 \pm 6.3$ & $100.0 \pm 6.7$ &
        $100 \pm 15$ & $100 \pm 20$ 
\\
$[{\rm O\;III}]\;4959$ &
        $48.8 \pm 3.7$ & $41.5 \pm 6.0$ &
        $89.6 \pm 5.3$ & $77.9 \pm 9.9$ &
        $89 \pm 12$ & $59 \pm 20$ 
\\
$[{\rm O\;III}]\;5007$ &
        $145.5 \pm 8.6$ & $123 \pm 16$ &
        $257 \pm 14$ & $222 \pm 27$ &
        $213 \pm 18$ & $140 \pm 39$ 
\\
He I 5876 &
        \nodata & \nodata &
        $5.9 \pm 2.5$ & $4.7 \pm 2.3$ &
        \nodata & \nodata 
\\
$[{\rm O\;I}]\;6300$ & 
        \nodata & \nodata &
        $11.5 \pm 3.1$ & $8.9 \pm 3.0$ &
        \nodata & \nodata
\\
$[{\rm N\;II}]\;6548$ &
        $16.2 \pm 4.7$ & $11.1 \pm 4.0$ &
        $9.2 \pm 4.0$ & $7.0 \pm 3.5$ &
        \nodata & \nodata 
\\
${\rm H}\alpha\;6563$ &
        $402 \pm 23$ & $286 \pm 36$ &
        $366 \pm 21$ & $286 \pm 36$ &
        $487 \pm 31$ & $286 \pm 75$ 
\\
$[\rm{N\;II}]\;6583$ &
        $56.4 \pm 5.5$ & $38.9 \pm 6.5$ &
        $30.0 \pm 4.3$ & $22.8 \pm 4.8$ &
        $62 \pm 13$ & $34 \pm 14$
\\
He I 6678 & 
        \nodata & \nodata &
        $4.9 \pm 1.7$ & $3.7 \pm 1.5$ & 
        \nodata & \nodata 
\\
$[{\rm S\;II}]\;6716$ &
        $82.9 \pm 5.5$ & $56.2 \pm 7.7$ &
        $69.2 \pm 4.3$ & $52.2 \pm 6.8$ &
        $135 \pm 13$ & $73 \pm 22$ 
\\
$[{\rm S\;II}]\;6731$ &
        $60.4 \pm 4.5$ & $40.9 \pm 5.9$ &
        $48.7 \pm 3.3$ & $36.8 \pm 4.9$ &
        $93 \pm 12$ & $50 \pm 16$ 
\\
$[{\rm Ar\;III}]\;7136$ &
        \nodata & \nodata &
        $11.2 \pm 1.9$ & $8.3 \pm 1.9$ &
        \nodata & \nodata 
\\
\tableline
\end{tabular}
\tablenotetext{a}{
VCC 1554-6 and VCC 1554-7: The recession velocity of emission lines
for each object is $\langle v_{\odot} \rangle /c \simeq +0.093$,
which puts these two objects as galaxies in the background of the
Virgo Cluster.   
Because \halpha\ was not detected in VCC~1554-6, the observed line
ratios are not corrected and the spectrum is not considered in
subsequent analyses.
}
\tablecomments{
Same as Table~\ref{table_allvirgo_data1}.
}
\end{center}
\end{table}

% \clearpage

\begin{deluxetable}{ccccccccccc}
\rotate
\renewcommand{\arraystretch}{1.}
\tablecolumns{11}
\tabletypesize{\tiny} % small 11pt; footnotesize 10pt; scriptsize 8pt
\tablewidth{0pt}
\tablecaption{
Derived properties for \hii\ regions in Virgo dwarfs.
\label{table_vccspecprops}
}
\tablehead{
\colhead{VCC dI} & \colhead{$I$(H$\beta$)} & \colhead{$E(B-V)$} & 
\colhead{Adopted} & \colhead{$W_{\rm e}$(\hbeta)} & 
\colhead{$W_{\rm abs}$(\hbeta)} & \colhead{$n_e$} &
\colhead{$T_e({\rm O}^{+2})$} &
\multicolumn{2}{c}{12$+$log(O/H)} & \colhead{log(N/O)} \\
\cline{9-10}
\colhead{H II \#} & \colhead{(ergs s$^{-1}$ cm$^{-2}$)} &
\colhead{(mag)} & \colhead{$E(B-V)$} & \colhead{(\AA)} &
\colhead{(\AA)} & \colhead{(cm$^{-3}$)} & \colhead{(K)} &
\colhead{\othreea} & \colhead{Bright-line} & \\  
\colhead{(1)} & \colhead{(2)} & \colhead{(3)} & \colhead{(4)} & 
\colhead{(5)} & \colhead{(6)} & \colhead{(7)} & \colhead{(8)} &
\colhead{(9)} & \colhead{(10)} & \colhead{(11)}
}
\startdata
0512-3 & $(4.50 \pm 0.39) \times 10^{-16}$ & 
	$-0.10 \pm 0.14$ & 0 & $89 \pm 15$ & 1.59 & 68 & 
	\nodata & \nodata & $8.25 \pm 0.19$ & $< -1.85$ \\
% Lower-branch Pily: 8.01 
0512-4 & $(2.91 \pm 0.45) \times 10^{-16}$ & 
	$-0.12 \pm 0.23$ & 0 & $26.4 \pm 4.2$ & 1.59 & 100 & 
	\nodata & \nodata & $8.20 \pm 0.25$ & $-1.53 \pm 0.30$ \\
% Lower-branch Pily: 8.29 
0848-1 & $(1.06 \pm 0.13) \times 10^{-14}$ &
        $+0.048 \pm 0.022$ & 0.05 & $33.69 \pm 0.57$ & 1.59 & 100 &
        $12700 \pm 1000$ & $7.98 \pm 0.04$ & $8.07 \pm 0.12$ & 
        $-1.25 \pm 0.11$ \\
% Lower-branch Pily: 7.92
0888-1 & $(1.64 \pm 0.83) \times 10^{-15}$ & 
	$+0.17 \pm 0.13$ & 0.17 & $9.01 \pm 0.33$ & $1.38 \pm 0.52$ &
	103 & \nodata & \nodata & $8.10 \pm 0.18$ & $-1.31 \pm 0.06$ \\ 
% Lower-branch Pily: 8.39? (log O32 -ve)
0888-2 & $(2.1 \pm 1.3) \times 10^{-15}$ & 
	$+0.27 \pm 0.15$ & 0.27 & $6.59 \pm 0.31$ & $1.19 \pm 0.49$ &
	100 & \nodata & \nodata & $8.27 \pm 0.15$ & $-1.54 \pm 0.12$ \\
% Lower-branch Pily: 8.37? (log O32 -ve)
1114-1 & $(3.70 \pm 0.35) \times 10^{-16}$ & 
	$-0.03 \pm 0.15$ & 0 & $47.0 \pm 5.9$ & 1.59 & 48 & 
	\nodata & \nodata & $8.10 \pm 0.30$ & $-1.20 \pm 0.21$ \\
% Lower-branch Pily: 8.55? (log O32 -ve)
1114-2 & $(4.36 \pm 0.41) \times 10^{-16}$ & 
	$-0.15 \pm 0.15$ & 0 & $260 \pm 130$ & 1.59 & 100 & 
	\nodata & \nodata & $7.97 \pm 0.27$ & $-1.21 \pm 0.25$ \\
% Lower-branch Pily: ??? (log O32 -ve)
1179-1 & $(3.15 \pm 0.69) \times 10^{-15}$ & 
	$+0.37 \pm 0.28$ & 0.37 & $1.61 \pm 0.23$ & $2.12 \pm 0.31$ & 
	135 & \nodata & \nodata & $8.33 \pm 0.30$ & $-1.52 \pm 0.16$ \\
% Lower-branch Pily: 8.47? (log O32 -ve)
1179-2 & $(3.27 \pm 0.26) \times 10^{-15}$ & 
	$+0.19 \pm 0.14$ & 0.19 & $6.09 \pm 0.28$ & $1.72 \pm 0.43$ & 
	100 & \nodata & \nodata & $8.26 \pm 0.16$ & $-1.40 \pm 0.07$ \\
% Lower-branch Pily: ??? (log O32 -ve)
1179-3 & $(5.53 \pm 0.42) \times 10^{-16}$ & 
	$-0.06 \pm 0.14$ & 0 & $13.79 \pm 0.66$ & 1.59 & 100 & 
	\nodata & \nodata & $7.95 \pm 0.22$ & $-1.12 \pm 0.13$ \\
% Lower-branch Pily: 7.83 
1179-4 & $(3.55 \pm 0.47) \times 10^{-16}$ & 
	$+0.07 \pm 0.20$ & 0.07 & $16.0 \pm 0.19$ & 1.59 & 100 & 
	\nodata & \nodata & $8.05 \pm 0.32$ & $-1.23 \pm 0.19$ \\
% Lower-branch Pily: 8.08
1200-2 & $(1.41 \pm 0.11) \times 10^{-15}$ & 
	$-0.15 \pm 0.13$ & 0 & $6.85 \pm 0.43$ & $2.68 \pm 0.66$ & 
	100 & \nodata & \nodata & $7.80 \pm 0.14$ & $-1.33 \pm 0.18$ \\
% Lower-branch Pily: 8.00
1249-1\tablenotemark{a} & $(3.68 \pm 0.70) \times 10^{-16}$ &
	0 & 0 & $26.8 \pm 0.59$ & 1.59 & 100 & 
	\nodata & \nodata & $8.32 \pm 0.20$ & $< -2.05$ \\
% Lower-branch Pily: (log O32 -ve)
1554-1 & $(1.86 \pm 0.68) \times 10^{-14}$ & 
        $-0.037 \pm 0.095$ & 0 & $112.3 \pm 6.8$ & 1.59 & 100 &
        $10100 \pm 2000$\tablenotemark{b} & 
        $8.38 \pm 0.04$ & $8.26 \pm 0.12$ & 
        $-1.63 \pm 0.10$\tablenotemark{b} \\
% Lower-branch Pily: 8.30
1554-2 & $(4.9 \pm 2.0) \times 10^{-16}$ & 
	$+0.09 \pm 0.10$ & 0.09 & \nodata\tablenotemark{c} & 1.59 &
	267 & \nodata & \nodata & $8.15 \pm 0.21$ & $-1.35 \pm 0.07$ \\
% Lower-branch Pily: 8.25
1554-3 & $(2.07 \pm 0.92) \times 10^{-15}$ & 
	$-0.06 \pm 0.11$ & 0 & \nodata\tablenotemark{c} & 1.59 & 
	100 & \nodata & \nodata & $8.04 \pm 0.21$ & $-1.49 \pm 0.16$ \\
% Lower-branch Pily: 7.99
1554-4 & $(4.8 \pm 3.3) \times 10^{-15}$ & 
	$+0.29 \pm 0.16$ & 0.29 & $4.33 \pm 0.33$ & 1.59 & 56 & 
	\nodata & \nodata & $8.35 \pm 0.34$ & $-1.50 \pm 0.11$ \\
% Lower-branch Pily: log O32 -ve
1554-5 & $(7.4 \pm 4.4) \times 10^{-16}$ & 
	$-0.12 \pm 0.14$ & 0 & \nodata\tablenotemark{c} & 1.59 & 
	100 & \nodata & \nodata & $8.28 \pm 0.24$ & $-1.59 \pm 0.10$ \\
% Lower-branch Pily: log O32 -ve
1554-7\tablenotemark{d} & $(2.64 \pm 0.70) \times 10^{-16}$ &
        $+0.11 \pm 0.34$ & 0.11 & $3.06 \pm 0.56$ & 1.59 &
        100 & \nodata & \nodata & 
        $\approx$ 8.85\tablenotemark{e} & $\approx -1.25$ \\
% O/H from formula between O/H and log(I(6583)/I(Ha))
% Get Te from plot of O/H vs Te from McGaugh 1991 ... Te about 6000K 
% Use expression to get N/O from Te and I(6583)/I(3727)
% logno = log(2.03*(tnorm)^(-0.02)*exp(-1.67/tnorm)*n2o2ratio) = -1.25
% [N II]/Ha calibs by van Zee / Denicolo : 9.14 / 8.86
1585-1 & $(6.0 \pm 4.9) \times 10^{-16}$ & 
	$+0.10 \pm 0.19$ & 0.10 & $66 \pm 13$ & 1.59 & 100 & 
	\nodata & \nodata & $8.07 \pm 0.32$ & $-1.11 \pm 0.07$ \\
% Lower-branch Pily: log O32 -ve
1789-1 & $(3.04 \pm 0.34) \times 10^{-15}$ & 
	$+0.23 \pm 0.18$ & 0.23 & $6.18 \pm 0.18$ & $0.94 \pm 0.29$ & 
	49 & \nodata & \nodata & $8.24 \pm 0.20$ & $-1.41 \pm 0.08$ \\
% Lower-branch Pily: log O32 -ve
2037-1 & $(2.67 \pm 0.18) \times 10^{-15}$ &
	$+0.15 \pm 0.13$ & 0.15 & $15.03 \pm 0.38$ & $1.24 \pm 0.62$ & 
	100 & \nodata & \nodata & $8.22 \pm 0.18$ & $-1.53 \pm 0.07$ \\
% Lower-branch Pily: 8.28; log O32 -ve
2037-2 & $(6.1 \pm 1.2) \times 10^{-16}$ & 
	$+0.22 \pm 0.28$ & 0.22 & $3.85 \pm 0.47$ & $1.48 \pm 0.42$ & 
	100 & \nodata & \nodata & $8.21 \pm 0.27$ & $-1.50 \pm 0.32$ \\
% Lower-branch Pily: (8.17); low O32 -ve
%
\enddata
\tablenotetext{a}{
The \hii\ region VCC 1249-1 is labelled as LR1 in \citet{lrm00}.
}
\tablenotetext{b}{
From the $I$(\ntwob)/$I$(\ntwotemp) intensity ratio, 
the derived [N~II] temperature is $13300 \pm 3000$~K.
With this [N~II] temperature, the log(N/O) value is $-1.46 \pm 0.10$,
which is 1.7$\sigma$ lower than the value listed in col. (11).
}
\tablenotetext{c}{
Negative equivalent width, owing to negative continuum.
}
\tablenotetext{d}{
Spectrum of background galaxy with 
$\langle v_{\odot} \rangle /c \simeq +0.093$.
}
\tablenotetext{e}{
In the absence of bright [O~III], the oxygen abundance was computed
using $\log\,[I(\ntwob)/I(\halpha)]$.
% \cite{ms02}
Expressions given by \cite{vanzee98sp} and \cite{denicolo02} 
yield similar results.
}
\tablecomments{
Column (1): H~II region.
Column (2): \hbeta\ intensity corrected for
underlying absorption and the adopted reddening.
Columns (3) and (4): Derived and adopted values of the reddening
from $F(\halpha)/F(\hbeta)$.
Column (5): Observed \hbeta\ emission equivalent width.
Column (6): Equivalent width of underlying absorption at \hbeta.
Column (7): Electron density.
Column (8): O$^{+2}$ temperature.
Column (9) and (10): Oxygen abundance derived from \othreea\ and 
with the bright-line method (McGaugh calibration), respectively.
Errors for \othreea\ abundances do not include errors for
reddening or temperature.
Column (11): Logarithm of the nitrogen to oxygen ratio.
}
\end{deluxetable}

% \clearpage

\begin{deluxetable}{ccccccccccc}
\rotate
\renewcommand{\arraystretch}{1.}  % 0.62, 0.95
\tablecolumns{11}
\tabletypesize{\tiny} % small 11pt; footnotesize 10pt; scriptsize 8pt
\tablewidth{0pt}
\tablecaption{
Stars, gas, and gas-deficiency indices for field and Virgo dwarfs.
\label{table_alldi_starsgas}
}
\tablehead{
\colhead{dI Name} & \colhead{$M_B$} & \colhead{log $M_{\rm H I}$} & 
\colhead{log $M_{\rm gas}$} & \colhead{log $M_{\rm H I}/L_B$} & 
\colhead{log $M_{\ast}$} & \colhead{$M_{\ast}/L_B$} & 
\colhead{Adopted} & \colhead{$\mu$} & \colhead{$\mu_p$} & 
\colhead{GDI} \\
& \colhead{(mag)} & \colhead{(\msun)} & \colhead{(\msun)} & 
\colhead{(\msun/\lsun)} & \colhead{(\msun)} & \colhead{(\msun/\lsun)} &
\colhead{12$+$log(O/H)} & & & \\
\colhead{(1)} & \colhead{(2)} & \colhead{(3)} & \colhead{(4)} & 
\colhead{(5)} & \colhead{(6)} & \colhead{(7)} & \colhead{(8)} & 
\colhead{(9)} & \colhead{(10)} & \colhead{(11)} 
}
\startdata
\multicolumn{11}{c}{{\sf Field dwarf irregulars}} \\
\tableline
DDO 187 & $-15.07$ & 8.25 & 8.39 & $+$0.032 & 8.01 & 0.61 & 
    7.69 & 0.706 & 0.770 & $+0.143$ \\  
GR 8 & $-12.19$ & 7.04 & 7.17 & $-$0.027 & 6.74 & 0.47 & 
    7.63 & 0.731 & 0.796 & $+0.156$ \\ 
Ho II & $-15.98$ & 8.93 & 9.06 & $+$0.344 & 8.54 & 0.90 & 
    7.76 & 0.771 & 0.735 & $-0.083$ \\
IC 10 & $-15.85$ & 8.14 & 8.27 & $-$0.394 & 8.62 & 1.21 & 
    8.20 & 0.312 & 0.431 & $+0.224$ \\
IC 1613 & $-14.53$ & 7.97 & 8.10 & $-$0.036 & 8.13 & 1.34 & 
    7.71 & 0.484 & 0.760 & $+0.529$ \\ 
IC 2574 & $-17.06$ & 9.16 & 9.30 & $+$0.146 & 8.95 & 0.86 & 
    8.09 & 0.690 & 0.520 & $-0.313$ \\ 
IC 4662 & $-15.84$ & 8.40 & 8.53 & $-$0.132 & 8.30 & 0.59 & 
    8.09 & 0.630 & 0.520 & $-0.196$ \\ 
Leo A & $-11.35$ & 6.89 & 7.02 & $+$0.154 & 6.36 & 0.43 & 
    7.36 & 0.820 & 0.884 & $+0.223$ \\
LMC & $-17.94$ & 8.82 & 8.96 & $-$0.544 & 9.37 & 1.01 & 
    8.35 & 0.279 & 0.306 & $+0.055$ \\
NGC 1560 & $-16.37$ & 8.85 & 8.98 & $+$0.107 & 8.74 & 0.99 & 
    $> 7.97$ & 0.638 & 0.608 & $-0.055$ \\
NGC 1569 & $-16.54$ & 7.80 & 7.93 & $-$1.008 & 8.58 & 0.59 & 
    8.19 & 0.185 & 0.440 & $+0.537$ \\
NGC 2366 & $-16.28$ & 8.95 & 9.08 & $+$0.243 & 8.79 & 1.22 & 
    7.91 & 0.662 & 0.648 & $-0.027$ \\
NGC 3109 & $-15.30$ & 8.94 & 9.07 & $+$0.624 & 8.31 & 1.00 & 
    7.74 & 0.851 & 0.746 & $-0.290$ \\
NGC 4214 & $-18.04$ & 9.24 & 9.37 & $-$0.169 & 9.40 & 0.98 & 
    8.24 & 0.485 & 0.398 & $-0.154$ \\ 
NGC 5408 & $-15.81$ & 8.25 & 8.38 & $-$0.269 & 8.51 & 0.98 & 
    8.01 & 0.429 & 0.580 & $+0.265$ \\
NGC 55 & $-18.28$ & 9.18 & 9.31 & $-$0.326 & 9.52 & 1.05 & 
    8.34 & 0.380 & 0.314 & $-0.127$ \\
NGC 6822 & $-14.95$ & 8.13 & 8.26 & $-$0.042 & 8.19 & 1.06 & 
    8.25 & 0.540 & 0.390 & $-0.264$ \\
Sextans A & $-14.04$ & 8.03 & 8.16 & $+$0.219 & 7.66 & 0.71 & 
    7.55 & 0.762 & 0.827 & $+0.173$ \\
Sextans B & $-14.02$ & 7.65 & 7.78 & $-$0.150 & 7.74 & 0.87 & 
    8.12 & 0.526 & 0.496 & $-0.051$ \\
SMC & $-16.56$ & 8.95 & 9.09 & $+$0.136 & 8.93 & 1.30 & 
    8.03 & 0.590 & 0.566 & $-0.044$ \\
UGC 6456 & $-13.90$ & 7.90 & 8.04 & $+$0.154 & 7.61 & 0.72 & 
    7.64 & 0.730 & 0.792 & $+0.148$ \\
WLM & $-13.92$ & 7.79 & 7.93 & $+$0.033 & 7.65 & 0.79 & 
    7.78 & 0.652 & 0.725 & $+0.149$ \\
\tableline
\multicolumn{11}{c}{{\sf Virgo dwarf irregulars}} \\
\tableline
VCC 0512 & $-15.82$ & 8.39 & 8.52 & $-$0.132 & 8.50 & 0.96 & 
    8.23 & 0.511 & 0.406 & $-0.183$ \\
VCC 0848 & $-16.39$ & 8.73 & 8.87 & $-$0.014 & 8.62 & 0.74 & 
    7.98 & 0.639 & 0.601 & $-0.071$ \\
VCC 0888 & $-16.12$ & 7.98 & 8.11 & $-$0.661 & 8.69 & 1.13 & 
%%  8.18 & 0.209 & 0.448 & $+0.487$ \\
    8.19 & 0.209 & 0.440 & $+0.473$ \\
VCC 1114 & $-16.46$ & 7.26 & 7.40 & $-$1.512 & 8.79 & 1.03 & 
%%  8.10 & 0.039 & 0.512 & $+1.409$ \\
    8.04 & 0.039 & 0.558 & $+1.489$ \\
VCC 1179 & $-16.25$ & 7.46 & 7.59 & $-$1.235 & 8.55 & 0.72 & 
%%  8.14 & 0.100 & 0.480 & $+0.922$ \\
    8.15 & 0.100 & 0.472 & $+0.908$ \\
VCC 1200 & $-16.34$ & 7.42 & 7.55 & $-$1.311 & 8.74 & 1.03 & 
    7.80 & 0.061 & 0.712 & $+1.585$ \\
VCC 1249 & $-16.86$ & 7.04\tablenotemark{a} & 7.18 & $-$1.893 & 8.98 &
    1.09 & 8.32 & 0.016 & 0.331 & $+1.491$ \\
VCC 1448 & $-17.19$ & 7.53 & 7.66 & $-$1.543 & 9.32 & 1.79 &
    8.22\tablenotemark{b} & 0.021 & 0.415 & $+1.510$ \\  % Z-MB
%   8.30\tablenotemark{b} & 0.021 & 0.347 & $+1.386$ \\  % Z-Mstar
VCC 1554 & $-18.97$ & 9.53 & 9.66 & $-$0.253 & 9.69 & 0.82 & 
    8.38 & 0.483 & 0.281 & $-0.379$ \\
VCC 1585 & $-15.92$ & 8.85 & 8.98 & $+$0.290 & 8.57 & 1.03 & 
    8.07 & 0.721 & 0.535 & $-0.351$ \\
VCC 1789 & $-16.05$ & 7.85 & 7.98 & $-$0.764 & 8.73 & 1.30 & 
    8.24 & 0.154 & 0.398 & $+0.562$ \\
VCC 2037 & $-15.32$ & 7.37 & 7.51 & $-$0.946 & 8.12 & 0.63 & 
%%  8.21 & 0.197 & 0.423 & $+0.476$ \\
    8.22 & 0.197 & 0.415 & $+0.461$ \\
\enddata
\tablenotetext{a}{
Measured \hi\ at the actual position of VCC~1249.
}
\tablenotetext{b}{
No \hii\ region spectrum was detected.
An estimate of the oxygen abundance was obtained from
the metallicity-luminosity relation (Equation~\ref{eqn_z_mb_fit}).
% using the derived value of $M_{\ast}$ and a fit between
% oxygen abundance and stellar mass \citep{lee01}.
}
\tablecomments{
Column (1): Name of the dI.
Column (2): Absolute magnitude in $B$.
Column (3): Logarithm of the \hi\ gas mass.
Column (4): Logarithm of the total gas mass.
Column (5): Logarithm of the \hi\ gas-to-blue-light ratio.
Column (6): Logarithm of the stellar mass.
Column (7): Stellar mass-to-light ratio in $B$.
Column (8): Adopted oxygen abundance.
Columns (9) and (10): Observed and predicted gas fractions,
respectively. 
Column (11): Gas-deficiency index (GDI).
For the list of field dIs, columns (1) to (9) inclusive 
are taken from \cite{lee03}.
}
\end{deluxetable}

\end{document}